\documentclass[usegraphicx, usenatbib]{mn2e}

\usepackage{color} \usepackage{url} \usepackage{amssymb}

\def\simleq{\; \raise0.3ex\hbox{$<$\kern-0.75em
\raise-1.1ex\hbox{$\sim$}}\; }

\font\japa = cmbx12 at 16.0truept \font\japb = cmr12 at 13.2truept

\begin{document}
\newcommand{\isnote}[1]{}

\title[SHADES survey design] {\japa The SCUBA HAlf Degree
Extragalactic Survey (SHADES) -- I.  Survey motivation, design and
data processing} \author[Mortier et al.]  {\japb
A.M.J. Mortier$^{1}$, S. Serjeant$^{1}$, J.S. Dunlop$^{2}$,
S.E. Scott$^{2}$, \newauthor\japb P. Ade$^{3}$,
D. Alexander$^{17}$, O. Almaini$^{4}$, \newauthor\japb
I. Aretxaga$^{5}$, C. Baugh$^{6}$, A.J. Benson$^{7}$, P.N. Best$^{2}$,
A.  Blain$^{8}$, J. Bock$^{9}$, C. Borys$^{8}$, \newauthor\japb
A. Bressan$^{10}$, C. Carilli$^{11}$, E.L. Chapin$^{5, 14}$,
S. Chapman$^{8}$, D.L. Clements$^{12}$, \newauthor\japb
K. Coppin$^{13}$, M. Crawford$^{2}$, M. Devlin$^{14}$,
S. Dicker$^{14}$, L.  Dunne$^{4}$, S.A. Eales$^{3}$, \newauthor\japb
A.C. Edge$^{6}$, D. Farrah$^{15}$, M. Fox$^{12}$, C. Frenk$^{6}$, E.
Gazta\~naga$^{5,16}$, W.K. Gear$^{3}$, \newauthor\japb
E. Gonzales-Solares$^{17}$, G.L. Granato$^{10}$, T.R. Greve$^{8}$,
J.A.  Grimes$^{7}$, J. Gundersen$^{18}$, \newauthor\japb
M. Halpern$^{13}$, P. Hargrave$^{3}$, D.H. Hughes$^{5}$,
R.J. Ivison$^{2, 19}$, M.J. Jarvis$^{7}$, T. Jenness$^{20}$,
\newauthor\japb R. Jimenez$^{14}$, E. van Kampen$^{2,21}$,
A. King$^{12}$, C. Lacey$^{6}$, A.  Lawrence$^{2}$, K. Lepage$^{13}$,
\newauthor\japb R.G. Mann$^{2}$, G. Marsden$^{13}$, P. Mauskopf$^{3}$,
B. Netterfield$^{22}$, S.  Oliver$^{23}$, L. Olmi$^{24}$,
\newauthor\japb M.J. Page$^{25}$, J.A. Peacock$^{2}$,
C.P. Pearson$^{26}$, W.J. Percival$^{2}$, A. Pope$^{13}$,
R.S. Priddey$^{27}$, \newauthor\japb S. Rawlings$^{7}$,
N. Roche$^{2}$, M. Rowan-Robinson$^{12}$, D. Scott$^{13}$, K.
Sekiguchi$^{28}$, M. Seigar$^{20,29}$, \newauthor\japb
L. Silva$^{30}$, C. Simpson$^{31}$, I. Smail$^{6}$,
J.A. Stevens$^{19}$, T.  Takagi$^{1}$, G. Tucker$^{32}$,
\newauthor\japb C. Vlahakis$^{3}$, I. Waddington$^{23}$,
J. Wagg$^{5}$, M. Watson$^{33}$, C.  Willott$^{34}$,
M. Vacarri$^{12}$\\ $^{1}$ Centre for Astrophysics and Planetary
Science, School of Physical Sciences, University of Kent, Canterbury,
Kent, CT2 7NR, UK\\ $^{2}$ Institute for Astronomy, University of
Edinburgh, Royal Observatory, Blackford Hill, Edinburgh, EH9 3HJ, UK\\
$^{3}$ Cardiff School of Physics and Astronomy, Cardiff University, 5,
The Parade, Cardiff, CF24 3YB, UK\\ $^{4}$ The School of Physics and
Astronomy, University of Nottingham, University Park, Nottingham, NG7
2RD, UK \\ $^{5}$ Instituto Nacional de Astrof\'{\i}sica, \'{O}ptica y
Electr\'{o}nica (INAOE), Apartado Postal 51 y 216, 72000 Puebla, Pue.,
Mexico\\ $^{6}$ Institute for Computational Cosmology, University of
Durham, South Rd, Durham DH1 3LE, UK\\ $^{7}$ Dept. of Astrophysics,
Denys Wilkinson Building, Keble Road, Oxford, OX1 3RH, UK\\ $^{8}$
Caltech, 1200 E. California Blvd, CA 91125-0001, USA\\ $^{9}$ Jet
Propulsion Laboratory, 4800 Oak Grove Drive, Pasadena, California
91109, USA\\$^{10}$ Osservatorio Astronomico di Padova, Vicolo
dell'Osservatorio, 5, I-35122, Padova, Italy\\ $^{11}$ National Radio
Astronomy Observatory, P.O. Box O, Socorro, NM 87801, U.S.A\\ $^{12}$
Astrophysics Group, Blackett Laboratory, Imperial College, Prince
Consort Rd., London SW7 2BW, UK\\
$^{13}$ Department of Physics \& Astronomy, University of British
Columbia, 6224 Agricultural Road, Vancouver, B.C., V6T 1Z1, Canada\\
$^{14}$ Department of Physics \& Astronomy, University of
Pennsylvania, 209 South 33rd Street, Philadelphia, PA 19104-6396,
USA\\ $^{15}$ Infrared Processing Analysis Center, California
Institute of Technology, Jet Propulsion Laboratory, Pasadena CA 91125,
USA\\ $^{16}$ Institut d'Estudis Espacials de Catalunya, IEEC/CSIC, c/
Gran Capita 2-4, 08034, Barcelona, Spain\\ $^{17}$ Institute of
Astronomy, University of Cambridge, Madingley Road, Cambridge CB3 0HA,
UK\\ $^{18}$ Experimental Cosmology Lab, Department of Physics,
University of Miami, 1320 Campo Sano Drive, Coral Gables, FL 33146\\
$^{19}$ UK ATC, Royal Observatory, Blackford Hill, Edinburgh, EH9 3HJ,
UK\\ $^{20}$ Joint Astronomy Centre, 660 N.\ A`oh\={o}k\={u} Place,
University Park, Hilo, Hawaii, 96720, USA \\ $^{21}$ Institute for
Astrophysics, University of Innsbruck, Technikerstr. 25, A-6020
Innsbruck, Austria\\ $^{22}$ Department of Astronomy and Astrophysics,
University of Toronto, 60 St.  George Street, Toronto, Ontario, M5S
3H8, Canada\\ $^{23}$ Astronomy Centre, University of Sussex, Falmer,
Brighton BN1 9QH, UK\\ $^{24}$ Osservatorio Astrofisico di Arcetri,
Largo E. Fermi 5, I-50125 Firenze, Italy\\ $^{25}$ Mullard Space
Science Laboratory (MSSL), University College London, Holmbury
St. Mary, Dorking, Surrey, RH5 6NT, UK\\ $^{26}$ Institute of Space
and Astronautical Science (ISAS), Yoshinodai 3-1-1, Sagamihara,
Kanagawa 229 8510, Japan\\ $^{27}$ Department of Physics, Astronomy \&
Mathematics, University of Hertfordshire, College Lane, Hatfield,
Hertfordshire AL10 9AB, UK\\ $^{28}$ Subaru Telescope, National
Astronomical Observatory of Japan, 650 North A'ohoku Place, Hilo, HI
96720, USA\\ $^{29}$ Department of Physics \& Astronomy, University of
California Irvine, 4129 Frederick Reines Hall, Irvine, CA 92697-4575,
USA\\ $^{30}$ Osservatorio Astronomico di Trieste, Via Tiepolo 11,
I-34131, Trieste, Italy\\ $^{31}$ Department of Physics, University of
Durham, South Road, Durham DH1 3LE, UK\\ $^{32}$ Department of
Physics, Brown University, 182 Hope Street, Box 1843, Providence, RI
02912, USA\\ $^{33}$ Department of Physics \& Astronomy, University of
Leicester, Leicester LE1 7RH, UK\\ $^{34}$ Herzberg Institute of
Astrophysics, National Research Council, 5071 West Saanich Rd,
Victoria, B.C., V9E 2E7, Canada } \date{in original form February
2004, revision 1.4 March 2005}

\maketitle

\begin{abstract}
The SCUBA HAlf Degree Extragalactic Survey (SHADES) is a major new
blank-field extragalactic sub-mm survey currently underway at the
James Clerk Maxwell telescope.  Ultimately, SHADES aims to cover half a
square degree at 450 and 850\,$\mu$m to a 4-$\sigma$ depth of $\simeq
8$\,mJy at 850\,$\mu$m.  Two fields are being observed, the
Subaru/XMM--{\it Newton\/} Deep Field (SXDF) ($02^{\rm h}18^{\rm
m}-05^\circ$) and the Lockman Hole East ($10^{\rm h}52^{\rm
m}+57^\circ$).  The survey has three main aims: i) to investigate
the population of high-redshift sub-mm galaxies and the cosmic history
of massive dust enshrouded star-formation activity, ii) to
investigate the clustering properties of sub-mm--selected galaxies in
order to determine whether these objects could be progenitors of
present-day massive ellipticals, and iii) to investigate the fraction
of sub-mm-selected sources that harbour active galactic nuclei.  To
achieve these aims requires that the sub-mm data be combined with
co-spatial information spanning the radio--to--X-ray frequency
range.  Accordingly SHADES has been designed to benefit from ultra-deep
radio imaging obtained with the VLA, deep mid-infrared observations from the
{\it Spitzer\/} Space Telescope, sub-mm mapping by the Balloon-borne
Large Area Sub-millimetre Telescope (BLAST), deep near-infrared
imaging with the UK Infrared Telescope, deep optical imaging with the
Subaru telescope, and deep X-ray observations with the XMM--{\it
Newton\/} observatory.  It is expected that the resulting extensive
multi-wavelength dataset will provide complete photometric redshift
information accurate to $\delta z \stackrel{<}{_\sim} 0.5$, as well as
detailed spectral energy distributions for the vast majority of the
sub-mm-selected sources.  In this paper, the first of a series on
SHADES, we present an overview of the motivation for the survey,
describe the SHADES survey strategy, provide a detailed
description of the primary data analysis pipeline, and demonstrate the
superiority of our adopted matched-filter source extraction technique
over, for example, Emerson-II style methods.  We also report on the
progress of the survey. As of February 2004, 720\, arcmin$^2$ had been mapped
with SCUBA (about 40\% of the anticipated final total area) to a
median $1\, \sigma$ depth of 2.2\,mJy per beam at 850\,$\mu$m (25\,mJy
per beam at 450\,$\mu$m), and the source extraction routines give a
source density of $650\pm 50$\, sources deg$^{-2}$ $>3\, \sigma$ at
850\,$\mu$m.  Although uncorrected for Eddington bias, this source
density is more than sufficient for providing enough sources to answer
the science goals of SHADES once half a square degree is observed.  A
refined re-analysis of the original 8-mJy survey Lockman hole data was
carried out in order to evaluate the new data reduction pipeline.  Of
the 17 most secure sources in the original sample, 12 have been
re-confirmed, including ten of the eleven for which radio identifications 
were previously secured.

\end{abstract}


\section{Introduction}
Theories of galaxy formation and evolution, embedded within
hierarchical structure formation models, can describe many of the
observed features of galaxies \citep{Cole, Granato00, Hatton}.  While
local galaxies can in some cases still provide constraints on the high
redshift populations \citep{Panter, Heavens}, the bulk of the
constraints on models of galaxy evolution come either from the
integral constraint from the far-infrared background \citep[e.g.][and
references therein]{Dwek, Gispert}, or directly from high-redshift
galaxy surveys \citep[e.g.][]{steidel}, the most ground-breaking of
which were the Canada-France Redshift Survey \citep{Lilly} and Hubble
Deep Field North \citep[HDF-N:][]{Williams}.

Such optical surveys have led to a great deal of progress in
understanding the assembly of stellar populations, and hierarchical
galaxy formation models are in increasingly good agreement with many
(but not all) of these observations \citep[e.g.][]{Cole, Somerville,
vanKampen99, Kauffmann, Guiderdoni, Blain99b}.

However, the discovery of a substantial population of faint sub-mm
galaxies \citep{Smail97, Hughes98,Barger98} has posed serious problems for
the current generation of galaxy-formation models 
based on hierarchical structure growth. Models of the optical/UV
spectral energy distributions (SEDs) of the galaxy population in the
HDF-N \citep{Thompson} predict only sub-mJy/$\mu$Jy-level 850\,$\mu$m
flux densities but the sub-mm point sources in the HDF-N field have
850\,$\mu$m flux densities of several mJy \citep{Hughes98, Serjeant03a,
Borys}.  This shows that there is a population of star forming galaxies
that are heavily obscured by dust and have much higher infrared
luminosities than would be inferred from the optical/UV observations
alone. If these galaxies are at high redshifts (as current data imply),
and if their emission is powered by star formation with a standard
solar neighbourhood initial mass function (IMF), then their observed
850\,$\mu$m flux-densities of several mJy imply star formation rates
$\sim 1000\,{\rm M_{\odot}yr^{-1}}$. Moreover,
the far-infrared (FIR) luminosity density implied by the sub-mm galaxy
population suggests that these infrared-luminous galaxies contributed
several tens of percent of the volume-averaged star formation density
at $z \simeq 2$ \citep[e.g.][]{Smail97, Barger99}.

  One attractive interpretation of
the sub-mm galaxy population is that these violently star-forming
galaxies are progenitors of present-day massive ellipticals
\citep{Hughes98, Scott02}.  There are four main pieces of evidence in
support of this \citep[e.g.]{Dunlop02}.  Firstly, the star formation rates
inferred from the sub-mm flux densities are sufficient to construct
the stellar population of even the most massive elliptical galaxy in
$\sim1$\,Gyr; secondly, the $K$-band morphologies of sub-mm-selected
galaxies resemble those of radio galaxies which locally are hosted in
giant ellipticals \citep[e.g.]{Lutz}; thirdly, the comoving number density
of bright sub-mm sources in the redshift interval $z \simeq 2$--$3$ is
comparable with the present-day number density of bright $>2$--$3
L^{\star}$ ellipticals \citep{Scott02}; and fourthly, tentative
detections of clustering suggest that the sub-mm galaxies trace the
collapse of rare, high density over-densities at high redshift
\citep{Almaini,Greve,Blain04b}.  Furthermore, the high
dynamical masses suggested by CO observations imply massive systems,
and the gas masses implied by the CO luminosities suggest extensive
star formation \citep{Genzel}; also, the dynamical, gas and stellar
masses estimated in the rest-frame optical/UV for these galaxies
indicate that they are both massive, gas-rich and already contain
significant stellar population \citep{Swinbank04,Smail04}.

A complication to this interpretation is that theory suggests a less
direct relation between local galaxies and their high-redshift
antecedents.  CDM simulations inevitably predict that massive galaxies
today are assembled hierarchically from a large number of smaller
fragments that existed at high redshift.  Conversely, the majority of
the rare early-collapsing objects at $z\simeq 2$--$3$ should be found
inside massive galaxies at the present \citep{Baugh98}.  If this is
true, then it suggests a more detailed set of questions that new, larger sub-mm surveys should attempt to settle:
(1) what fraction of present-day massive ellipticals have merged with
at least
one SCUBA galaxy?; (2) what fraction of SCUBA galaxies will end up in
a present-day massive elliptical? (3) if the answer to the second
question is close to 100\%, how close are the SCUBA galaxies to the
end of the merger process? There are claims \citep{Bell04} that the
total mass of stars in ellipticals has roughly doubled since redshift
$z = 1$.  If this increase applies to the most luminous ellipticals,
this would mean that most present ellipticals were only in
the earliest phases of assembly at $z=2$--$3$.  There is thus some
uncertainty over whether a given SCUBA galaxy represents a late phase
in the construction of an elliptical, or an early phase in the
assembly of its nucleus.  Uncertainties in the true star-formation rate
and the possible lifetime of the starburst contribute to this
uncertainty, which can only be addressed statistically by
looking at the population as a whole.

Detailed models of the hierarchical assembly of galaxies, based on
standard prescriptions for gas cooling, star formation and feedback,
in general do not predict large numbers of SCUBA galaxies with
star-formation rates $\sim 1000 \rm{M_{\odot}} {\rm yr}^{-1}$.  However,
\citet{Baugh} have shown that the observed number counts and redshift
distributions of the sub-mm sources can be reproduced in CDM models if
the star formation in these objects is occurring with a top-heavy IMF
\citep[e.g.][]{Larson}, implying that the same sub-mm flux can be
produced with star formation rates $\sim 5$ times lower than for a
Salpeter IMF.  An alternative model that can explain the sub-mm
galaxies within the framework of CDM is that of \citet{Granato04}, who
propose modifications to the treatments of gas cooling and feedback,
as opposed to modifications to the IMF.  Finally, the treatment of
virialization and the survival of subhaloes in the model of
\citet{vanKampen99} produces star formation histories which allow for
much higher star formation rates at early times (especially in
bursts), and therefore predicts sufficient numbers of sub-mm galaxies
at high redshifts \citep{vanKampen03}.  SHADES will provide the means
to distinguish between these and other alternatives \citep{vanKampen}.

Of course, the radical possibility that the hierarchical orthodoxy may
be flawed in some way is worth keeping open, in order that the
standard theory can be properly verified.  Furthermore, even the
interpretation of all SCUBA sources as dusty starbursts could still be
subject to future revision.  Although the majority of the bright
sub-mm sources have secure extragalactic identifications, and in many
cases redshifts, 
it remains possible (albeit increasingly unlikely) that a
fraction of sub-mm objects could be Galactic in origin \citep{Sciama,
Lawrence}.  Even accepting that the sources are extragalactic, the
starburst model is not unchallenged: \citet{Efstathiou} and
\citet{Kaviani} have successfully modelled the emission of some sub-mm
sources as extended cirrus in galaxies heated by the interstellar
radiation field.  This is possible partly because the far-infrared
colour temperatures are not well-constrained with existing data.  This
interpretation is supported by observational evidence from
\citet{Farrah04}, using sub-mm detections of a galaxy at $z\simeq 0.5$
that hosts a Type Ia supernova.  The rest-frame optical luminosities
and colours of sub-mm galaxies require such cirrus galaxies to be more
heavily extinguished than their local counterparts -- Efstathiou \&
Rowan-Robinson finding a best fit extinction co-efficient of
$A_V\simeq 1$--$3$ for the high redshift galaxies as opposed to
$A_V\simeq
0.4$--$0.9$
for the local galaxies in their sample.  Such an interpretation may be
testable using high-resolution observations with the Plateau de Bure 
Interfermometer, and will be conclusively answered using
the Atacama Large Millimeter Array (ALMA) with a
spatial resolution of $\stackrel{<}{_\sim} 0.1^{\prime\prime}$.

At present, the determination of the redshift distribution and
clustering of the sub-mm galaxy population offers the best
available method to constrain the properties of the over-densities
hosting bright sub-mm galaxies, and hence to differentiate between
alternative models of galaxy evolution \citep{vanKampen}.  To this
end, the SCUBA HAlf Degree Extragalactic Survey (SHADES,
\url{http://www.roe.ac.uk/ifa/shades/}) consortium is mapping $0.5$\,
deg$^2$ with the Sub-millimetre Common User Bolometer Array (SCUBA) at
the James Clerk Maxwell Telescope (JCMT) on Mauna Kea, Hawaii.  SHADES
expects to produce a complete sample of $\sim300$ bright sub-mm
sources at 850\,$\mu$m.  The survey is being carried out in fields with
abundant supporting multi-wavelength data (see Section
\ref{sec:fields} for full details).

In this paper, the first in a series of papers on SHADES, we present
an overview of the motivation for the survey, discuss the adopted
observing strategy, and describe the operation of, and first results
from the primary SHADES data reduction pipeline.  However, we stress
that three additional and independent reductions of the SHADES data
are currently underway within the consortium, and that we aim to take
advantage of these multiple reductions to maximise the robustness of
the final 850\,$\mu$m source list.  These alternative reductions, and
the outcome of cross-referencing the resulting maps and source lists
will be presented elsewhere. 
The current paper is structured as follows.  Section \ref{sec:motive}
describes the aims of the SHADES survey in more detail, and Sections
\ref{sec:method} and \ref{sec:DataReduction} present the data
acquisition and analysis methods being used.  Section
\ref{sec:sourcextr} outlines the source-extraction methods under
development for this large survey and in Section \ref{sec:analysis} we
present the progress of the survey so far and Section
\ref{sec:discussion} concludes the paper. Throughout, we 
assume a cosmology with
$\Omega_{\rm M}=0.3$, $\Omega_\Lambda=0.7$ and a Hubble constant of
$H_0=72$ km s$^{-1}$ Mpc$^{-1}$.

\section{Motivation for the survey}\label{sec:motive}
\subsection{Background: The problem of cross-identifications}\label{sec:crossid}

The faint optical/near-infrared identifications of sub-mm galaxies
\citep[e.g.][]{Lilly99, Barger00, Ivison02, Smail02, Serjeant03a,
Serjeant03b, Webb03, Clements, Dunlop04, Wang04, Borys04} and the
broad ($\sim10$--$15^{\prime\prime}$) beams of the largest current
sub-mm/mm-wave telescopes, together present difficulties for the
unambiguous identification of sub-mm galaxies \citep{Hughes98}.

However, extensive long-term efforts towards identification made in
the radio \citep[e.g.][]{Ivison04}, supported in some cases by mm-wave
interferometry \citep[e.g.][]{Downes, Gear, Lutz}, have produced radio
identifications for $\sim$ 50-70\% of the brighter sub-mm sources.
These radio detections have been successfully exploited to derive
accurate (i.e.  sub-arcsec) positions for sub-mm galaxies, thus
facilitating further spectroscopic study.  As a result, 
spectroscopic redshifts for
approximately 90 radio-detected sub-mm sources have been published to
date \citep{Ivison98, Ivison00, Chapman03,Chapman05}.

The spectroscopic follow up is of course biased against those sources
at redshifts where no spectral features fall within the spectroscopic
range, most notably $1.2<z<1.7$, and may also exclude the highest
redshift objects (due to the less favorable K-correction in the radio
waveband compared to the sub-mm for redshifts $z
\stackrel{>}{_\sim}3$).  Together these effects result in moderate
incompleteness in the final redshift surveys \citep{Chapman05} even if
all identifications are robust.

Recently, rapid detections of SCUBA galaxies have been made with the
{\it Spitzer\/} Space Telescope, in integrations of only $\sim10$
minutes \citep{egami, Ivison04, Serjeant04, Frayer, Charmandaris}.
This small sample of identifications, in conjunction with the abundant
{\it Spitzer\/} coverage of our fields, shows the potential for
identification and follow-up of the sub-mm sources.  

Despite the radio and the {\it Spitzer\/} data, some sources may
remain unidentified.  Lack of a robust radio or mid-infrared 
indentification could have five
origins: (i) the sub-mm source could be spurious; (ii) the source
could be severely flux boosted \citep[][see Section
\ref{sec:Comp}]{Eddington}; (iii) the radio/far-IR emission could be
significantly more extended than the VLA synthesised beam; (iv) the
characteristic dust temperatures could be low; or (v) the source could
be at very high redshift.

Nevertheless, armed with lower resolution observations with the VLA,
along with the low-resolution sub-mm imaging to be provided by the
Balloon-borne Large Area Sub-millimetre Telescope \citep[BLAST,][see
Section \ref{sec:fields}]{Hughes02} and mid-infrared imaging from the
{\it Spitzer\/} Wide-Area Infrared Extragalactic Legacy Survey
\citep[SWIRE,][see Section \ref{sec:fields}]{Lonsdale03, Lonsdale04}
we anticipate being able to distinguish between these five
alternatives for most of the apparently unidentified sources.

The large positional uncertainty of the SCUBA sources may also lead to
unreliable identifications.  A measure of this is the
cross-identification limit, which we define to be one random source
per 10 search areas.  At this surface density the likelihood of a
spurious identification is given by $ p \simleq 1 - \exp^{- N(>S) \pi r^{2}}
\simeq 0.1 $, where $r$ is 7$\,^{\prime\prime}$ (the half
width half maximum of the JCMT beam at 850\,$\mu$m), and 
$N(>S)$ is the cumulative source counts of
other objects in the identification catalogue with fluxes greater than
$S$.  
For the {\it Spitzer\/} IRAC 3.6, 4.5\,$\mu$m bands, the resulting 
cross-identification
limits are 
$58$ and $60\,\mu$Jy respectively, well above the SWIRE
sensitivity and confusion limits (shown in Table \ref{tab:sirtf}).
For the IRAC 5.8 and 8.0\,$\mu$m and {\it Spitzer\/} MIPS 24$\mu$m
bands, the cross-identification 
limits are $48$, $44\,\mu$Jy and $120\,\mu$Jy,
similar to the SWIRE sensitivities but still well above the confusion
limit \citep{Fazio,Marleau}.  The 70 and 160\,$\mu$m cross-identification
limits
of $0.93$ and $9.1\,mJy$ are well below the confusion and SWIRE
sensitivity levels \citep{Dole}.  This highlights the need for careful
analyses to identify the SHADES galaxies at
wavelengths shorter than 24\,$\mu$m.  Our abundant multiwavelength
coverage, especially optical and near-infrared, can be used to find objects by
looking at outliers in colour-space \citep[e.g.]
{Smail02,Webb03,Pope}.

To illustrate the potential power of the
multi-frequency/multi-facility approach adopted for SHADES, we have
investigated the properties of four template galaxies, whose SEDs have
been normalised to the SHADES survey depth of $8$\,mJy at 850\,$\mu$m,
as the assumed redshift is varied.  Table \ref{tab:sirtf} shows the
flux densities expected from our four template galaxies.  Note the
similarity of the SWIRE limits to the Arp 220 model SED flux densities
and also to the depth required for reliable identifications.
Fig.\,\ref{fig:simSEDs} shows the four example model SEDs as the
assumed redshift is varied from $z = 0$ to $z = 4$.  The key points
illustrated by these models can be summarized as follows.

\noindent
i) The Arp 220 SED with the SHADES normalisation shows that BLAST
detections of Arp 220-like SEDs should be possible to $z \simeq 2-3$
and that detections in 3.6 and 4.5\,$\mu$m in conjunction with
non-detections at 5.8, 8 and 24\,$\mu$m can be used to differentiate
Arp 220-like SEDs from other forms.  The relative number counts of
\citet{huang} show that higher wavelength drop-outs do occur;
\citet{LeFloch}, using preliminary {\it Spitzer\/} MIPS (24\,$\mu$m)
results, suggest that the SEDs of high redshift ($z \geq 1$) sources
in the Lockman Hole East and the Extended Groth Strip are well fitted by
an Arp 220-like model.\\ 
ii) M82 is another prototypical star-forming galaxy and
galaxies with M82-like SEDs should be visible at BLAST depths out to
$z \simeq 3$ and {\it Spitzer\/} SWIRE depths out to $z \simeq 4$
using the IRAC (3.6-8\,$\mu$m) wavebands.\\ 
iii) The SED of HR10, a
high-redshift, extremely-red galaxy, is well matched out to redshifts
of $z \simeq 2-3$ with BLAST and $z \simeq 3-4$ at shorter
wavelengths.  HR10 would be defined as a ultraluminous infrared galaxy
(ULIRG) by its infrared luminosity.\\ 
iv) The SED of NGC 1068, dominated by an AGN-heated dusty torus, is
discussed in more depth in Section \ref{subsec:AGN}.

These models illustrate that the mid-infrared and far-infrared 
flux-density limits for the coverage of the SHADES fields
will detect sub-mm galaxies to redshifts $z\stackrel{<}{_\sim}4$,
comparable to that from the deepest radio integrations.  
We can confirm these expectations using the properties of 
existing SCUBA galaxies which have been detected 
at 24\,$\mu$m \citep{egami, Frayer}. These have been found to have 
24\,$\mu$m flux densities ranging from $80$ to $2800\, \mu$Jy, although the
median flux is $230\, \mu$Jy and half of the detections lie in the
range $160-370\, \mu$Jy. Such measurements 
compare favourably with the anticipated
SWIRE $5\, \sigma$ survey limit of $105\, \mu$Jy at 24\,$\mu$m
(Lonsdale et al. 2004).  The deeper {\it Spitzer\/} GTO data may also
detect many of the sub-mm galaxies at other wavelengths.

\subsection{SHADES science goals}
\subsubsection{Measurement of the cosmic history of massive dust-enshrouded 
star-formation activity} A key constraint in the phenomenology of
galaxy formation is the cosmic history of dust-enshrouded star
formation, observed via the evolution of far-infrared luminosity
density. The flat sub-mm selection function is well-known to give
sub-mm galaxy surveys an informative constraint on the far-infrared
luminosity density throughout most of the Hubble volume
\citep[e.g.][]{Franceschini, Blain96}.  Spectroscopic redshifts or
photometric redshift estimates are clearly essential to constrain the
evolving far-infrared luminosity density, but the identifications have
proved challenging (see above).  Moreover, even when secure
identifications are available, the optical or near-infrared
spectroscopic follow-up observations are time-consuming, typically
requiring $>2$ hours of integration on each target with 8--10m-class
telescopes, and are not guaranteed of success
\citep[e.g.][]{Chapman03, Simpson04}.

However, even in the absence of an optical or near-infrared
identification photometric redshifts can still be derived on the basis
of long-wavelength (i.e. radio--to--far-infrared) photometry.  The
simplest of these methods uses a single colour ratio between
observations at 850\,$\mu$m and 1.4\,GHz \citep{Carilli, Dunne00,
Rengarajan} to discriminate between low and high-$z$ star-forming
galaxies.  This method applied to a sample of 30 sources
\citep{Ivison02} produces estimated redshifts accurate to $\delta z
\simeq 0.4$--$1.5$.

Despite the dust temperature -- redshift degeneracy present in FIR
sources \citep{Blain99, Blain03}, extension of this method to include
multiple colour information in the radio to sub-mm bands allows the
derivation of photometric redshifts with accuracies of $\delta z
\simeq 0.5$ or better \citep{Yun, Hughes02, Aretxaga03, Aretxaga05,
Wiklind}, even taking into account the full range of dust properties
present in the local galaxy analogues.  The most recent comparison of
photometric and spectroscopic redshifts for blank field sub-mm sources
yield a dispersion of $\delta z \sim
0.3$ when three or more long-wavelength detections
are available \citep{Aretxaga05}.  Uncertainties due to the
temperature-redshift degeneracies could be further reduced by
performing spectroscopic calibration upon a representative subset of
the sample \citep{Blain04a}.

For those sources for which near-infrared or mid-infrared
identifications are secured, a complementary (and in principle
independent) method of estimating redshifts is to use the {\it
Spitzer\/} Space Telescope and corresponding ground based
near-infrared observations to identify the position of the redshifted
1.6\,$\mu$m peak of the near-infrared stellar continuum using the IRAC
3-8\,$\mu$m bands \citep{Simpson, Sawicki}.  This method gives an
indication of which objects lie at redshift $z \stackrel{>}{_\sim}
1.5$ and obtains redshift estimates accurate to $\delta z\sim0.5$ at
$z\stackrel{<}{_\sim}1.5$ independent of the BLAST and VLA
measurements.  Where optical photometry is available, the photometric
redshift is more accurate still \citep{Pope}.

Given the need to break the sample into a few redshift bins, while
maintaining statistically useful 
numbers in each bin, we require of the order of
100 sources in total in order for the $N(z)$ histograms to be able to
differentiate between available models \citep{vanKampen}. The
combined multiwavelength data are expected to yield a redshift resolution
of better than $\delta z \simeq 0.5$ for $300$ galaxies, using
spectroscopically calibrated photometric redshifts (Aretxaga et al. 2005).

\subsubsection{Determination of whether SCUBA galaxies are progenitors of 
present-day massive elliptical galaxies}

The high inferred star formation rates of sub-mm galaxies, while
consistent with expectations for proto-ellipticals, nevertheless do
not provide unambiguous evidence that they are progenitors of massive
elliptical galaxies.  This is both because the duration of the
starburst is unknown, and because the derived star-formation rate is
sensitive to the assumed temperature and initial mass function of the
mass distribution \citep{Larson, Baugh}.  In contrast, the clustering
of bright SCUBA sources on scales of up to $\sim10$\, Mpc offers a
potentially very powerful constraint on the nature of the sub-mm
population \citep{Percival}, given even a relatively broad ($\delta
z\simeq 0.5$) constraint on the estimated redshifts of individual
sources (van Kampen et al. 2005).  As discussed in van Kampen et
al. (2005), to sample the sub-mm population over such scales requires
a degree-scale survey ($1^\circ$ is $29$\,Mpc at $z=2$ in our adopted
cosmology).  Therefore, the desire to probe scales approaching 10\,Mpc
provides one of the primary motivations for the eventual areal
coverage goal of SHADES (i.e. half a square degree).

Previous blank sky surveys carried out with SCUBA have identified a
total of $\sim100$ sources in disparate areas on the sky.  These survey
fields include the Hubble Deep Field \citep{Hughes98, Serjeant03a,
Borys, Wang04, Borys04}, the Hawaii deep field regions
\citep{Barger98, Barger99}, the Lockman Hole East and the Elais N2
region in the 8-mJy Survey \citep{Scott02, Fox}, the {\it Spitzer\/}
northern continuous viewing zone \citep{Sawicki05}, and the
Canada-France Redshift Survey (CFRS) fields by the Canada UK Deep
Sub-millimeter Survey \citep[CUDSS; ][]{Eales99}.  However, the
resulting composite existing `sample' of sub-mm sources spans a wide
range in intrinsic luminosity and is distributed between many
small fields imaged to varying depths.  As a result it has proved
impossible to derive unambiguous constraints from the apparent
clustering strength of these sources \citep{Borys, Scott02}.  By
providing a complete and homogeneous sample of the most luminous
sub-mm sources in two wide-area fields, SHADES aims to provide the
first robust constraints of the clustering properties of the sub-mm
galaxy population.

\subsubsection{Determination of the fraction of SCUBA sources that harbour
obscured active galactic nuclei}\label{subsec:AGN} The {\it Spitzer\/}
$3.6$--$160\, \mu$m data from the SWIRE and GTO surveys
\citep{Lonsdale03, Lonsdale04, egami, huang} sample the spectral
energy distributions in the rest-frame near-infrared and
mid-infrared.  The latter is sensitive to the presence of AGN dust
tori, making the {\it Spitzer\/} data key to determining the AGN
bolometric fraction in these high-infrared-luminosity galaxies
\citep[e.g.][]{Almaini99, Efstathiou00, Farrah02}.  Most torus models
show warm colour temperatures which therefore would not contribute
significantly to the sub-mm flux.

X-ray visible AGN and sub-mm sources are only rarely coincident in
shallow X-ray observations \citep{Bautz, Fabian, Waskett, Almaini} and
yet seem to trace similar structures on arcminute scales in the Elais
N2 field \citep{Almaini}.  Possibly, these two populations represent
different, relatively short-lived, phases in the formation of massive
objects at high redshift. Alternatively, the
majority of SCUBA sources may contain a massive and active black hole
that is too heavily obscured to be detected with the current X-ray
surveys.  Such Compton-thick objects may be associated with the
formation of super-massive black holes.  Some X-ray objects have been
found to have a sub-mm source associated with them
\citep[e.g. ][]{Barger01} and \citet{Alexander,Alexander05} suggest
that even those sub-mm galaxies without an X-ray counterpart could
contain low luminosity AGN.  However, as Alexander et al. point out,
the sub-mm emission in these galaxies would be dominated by star
formation and not the AGN, so any estimates of star formation rates
would not be affected.

\begin{figure*}
\centering{ \includegraphics[width=8cm]{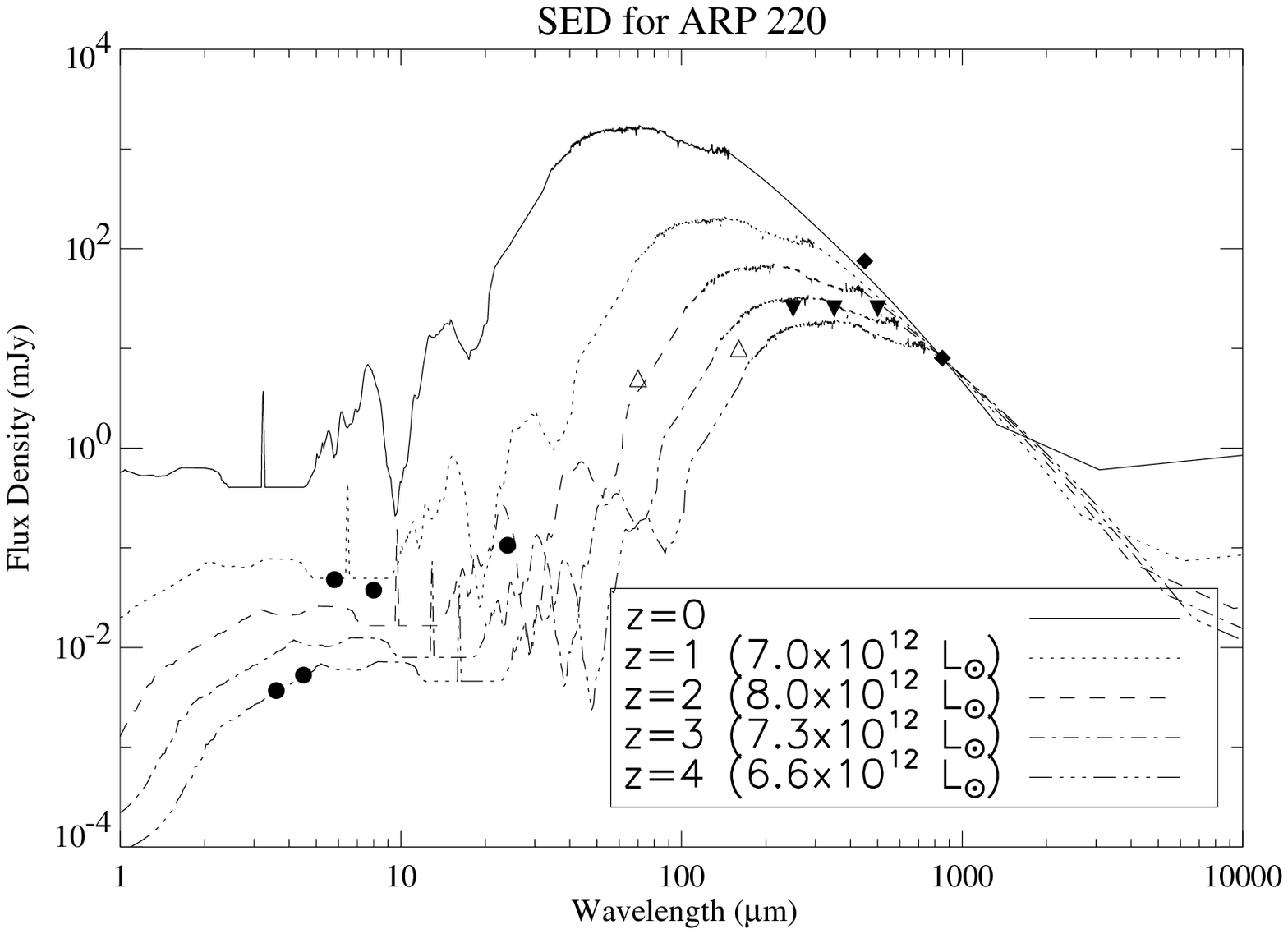}
\includegraphics[width=8cm]{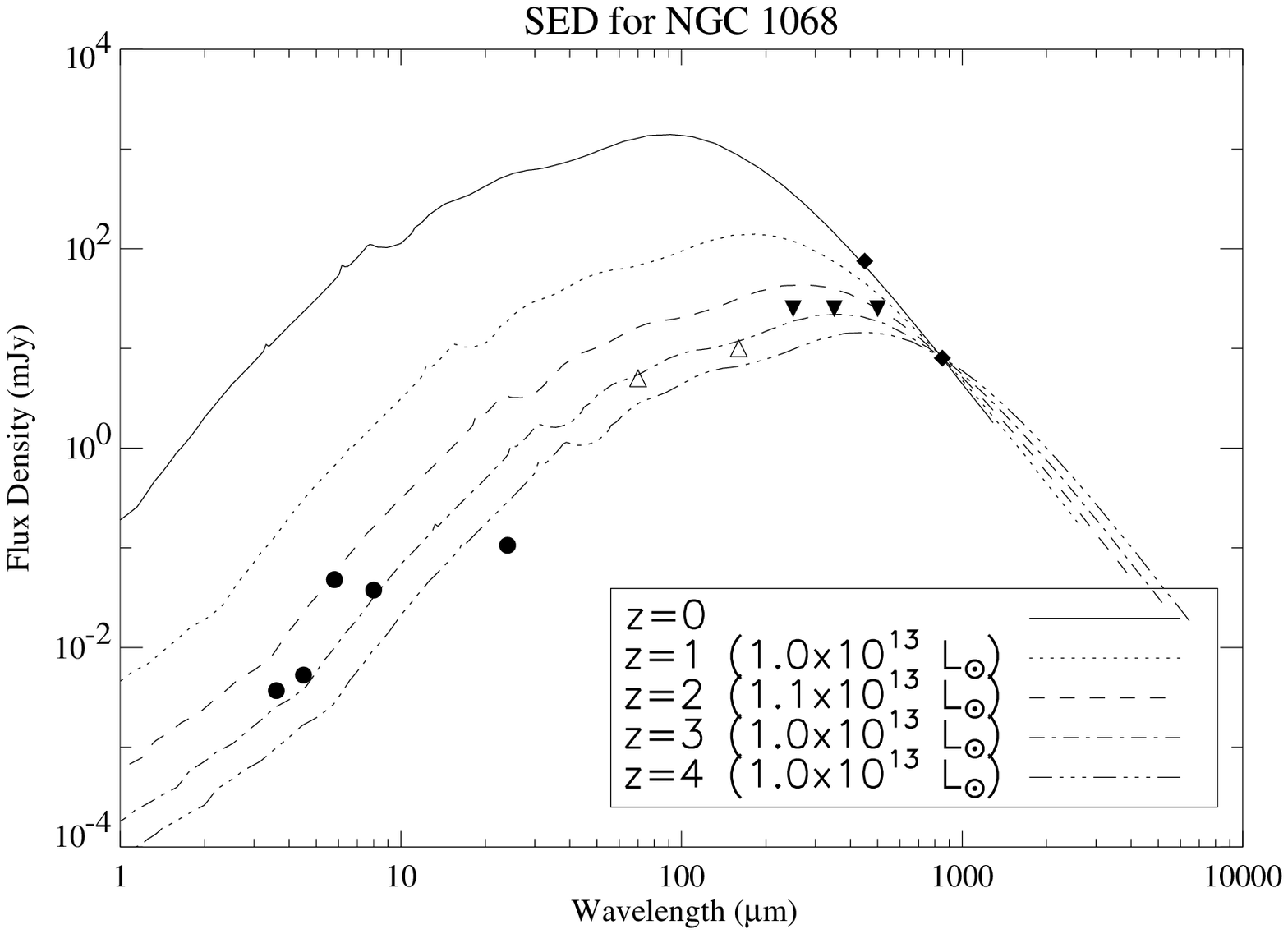}}
\includegraphics[width=8cm]{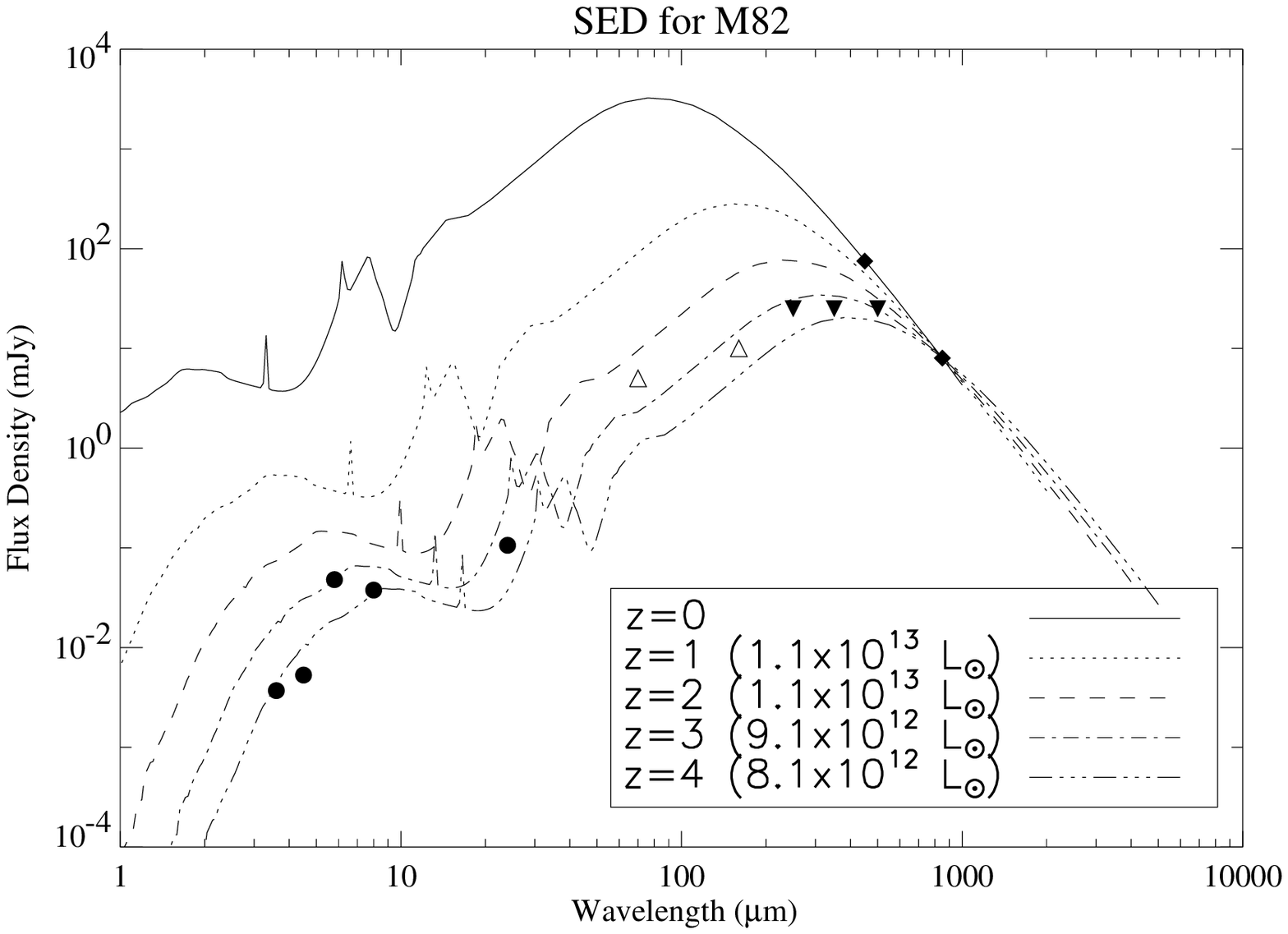}
\includegraphics[width=8cm]{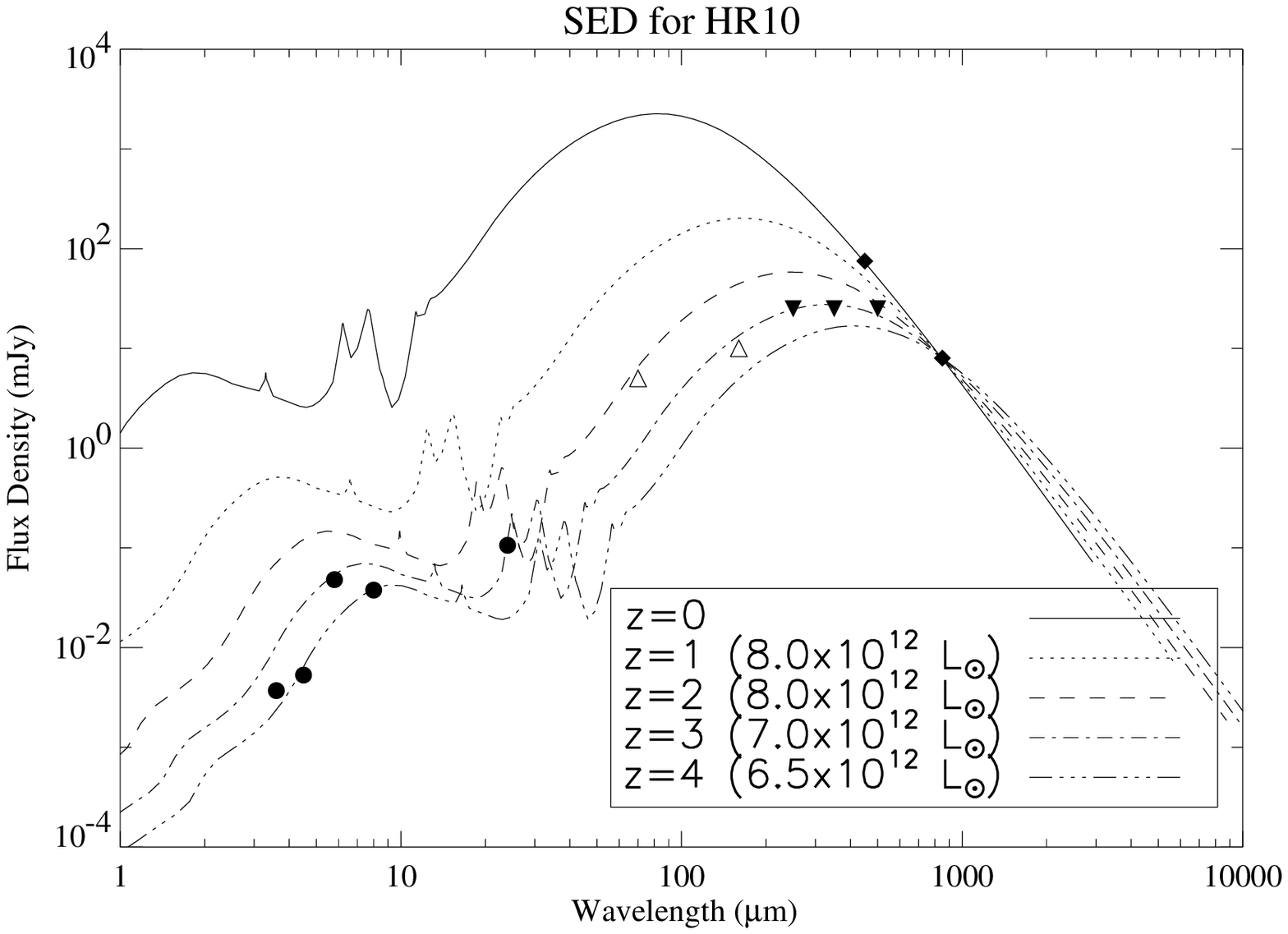}
\caption {Model spectral energy distribution plots of template
galaxies with survey sensitivities overplotted.  {\it Filled circles:}
SWIRE sensitivities 500\, seconds, 5\,$\sigma$ \citep{Lonsdale04}.
{\it Open triangles:} The pre-flight {\it Spitzer\/} sensitivities for
70 and 160\,$\mu$m are shown whilst the SWIRE team re-assess their
observing strategy due to the degraded MIPs sensitivities, although
currently these are expected to be two to three times worse.  {\it
Filled triangles:} The pre-flight BLAST confusion limit estimates
\citep[derived from ][]{Rowan-Robinson} of 25\,mJy; the sensitivity
estimates \citep{Hughes02} are in the range $15-25$\,mJy.  {\it Filled
diamonds:} SHADES survey sensitivities of 
8\,mJy at 850\,$\mu$m and 75\,mJy at
450\,$\mu$m.  Models have been normalised to the SHADES survey depth of
$8$\,mJy at 850\,$\mu$m.  {\it Top left panel:} Arp 220 -- a 
highly-obscured local starburst galaxy \citep[note: it is also a ULIRG
][]{Elbaz}.  {\it Top right panel:} NGC1068 -- a typical Seyfert
galaxy \citep{Efstathiou95}.  {\it Bottom left panel:} M82 -- an
irregular dusty star-forming galaxy \citep{Efstathiou00}.  {\it Bottom
right panel:} HR10 -- a high-redshift, extremely-red galaxy
\citep{Takagi}.  Bolometric (3--1000\,$\mu$m rest-frame) luminosities
for each model, as normalised, are shown in the legends.
\label{fig:simSEDs}}
\end{figure*}

The NGC 1068 SED in Fig.\,\ref{fig:simSEDs}, normalised to the SHADES
sensitivity, demonstrates that hyperluminous AGN dust tori in the
SHADES survey, with the AGN dominating even the sub-mm flux, should be
detectable in the {\it Spitzer\/} $24-160\, \mu$m and BLAST $250-500\,
\mu$m bands at redshifts of up to $z=2$.  In this extreme limiting
case, the SED is dominated by the dust torus, with little contribution
from circumnuclear star formation. \citet{Ivison04} and \citet{egami}
have already shown that {\it Spitzer\/} 24\,$\mu$m photometry is
efficient at demonstrating the presence of AGN in cases where the AGN
makes a much smaller bolometric contribution, compared to that of the
star formation; this AGN detection is particularly effective when
combined with photometry in the IRAC bands \citep{Ivison04}.
\renewcommand{\thefootnote}{\fnsymbol{footnote}}

\begin{table*}
\caption{\label{tab:sirtf} \textbf{Comparison of the {\it Spitzer\/}
survey flux density limits with galaxy populations normalised to the
SHADES depth of 8\,mJy at 850\,$\mu$m.  Also listed are the effective
cross-identification limits (as described in the text).}}
\begin{tabular}{llllllll}
\hline Wavelength & 3.6\,$\mu$m & 4.5\,$\mu$m & 5.8\,$\mu$m & 8.0\,$\mu$m
& 24\,$\mu$m & 70\,$\mu$m & 160\,$\mu$m\\ \hline
SWIRE $5\sigma$ in 500s \footnotemark[2] & $3.7\, \mu$Jy &
$5.3\,\mu$Jy & $48\,\mu$Jy & $37.7\,\mu$Jy & $106\,\mu$Jy &
$6$\,mJy\footnotemark[2] & $50$\,mJy\footnotemark[2]\\
$S_{conf}$\footnotemark[7]& $1.5\,\mu$Jy \footnotemark[5]& $
1.5\,\mu$Jy \footnotemark[5] & $8.3\,\mu$Jy\footnotemark[5] & $
5.4\,\mu$Jy\footnotemark[5] & $88\,\mu$Jy \footnotemark[6]&
$5.3\,$mJy\footnotemark[4] & $48\,$mJy\footnotemark[4] \\
$N_{src}$\footnotemark[3] & 1.7\footnotemark[5] &
1.4\footnotemark[5] & 0.84 \footnotemark[5] & 0.81\footnotemark[5] &
0.13\footnotemark[6] & 0.009\footnotemark[4] & 0.003\footnotemark[4]
\\ \hline
ARP 220 $z=2$ \footnotemark[1] & $22\, \mu$Jy & $23\, \mu$Jy & $26\,
\mu$Jy & $17\, \mu$Jy & $0.21$\,mJy & $3.7$\,mJy & $62$\,mJy \\
ARP 220 $z=4$ \footnotemark[1] & $3.5\, \mu$Jy & $5.2\, \mu$Jy &
$6.1\, \mu$Jy & $6.9\, \mu$Jy & $6.0\, \mu$Jy & $0.17$\,mJy & $4.3$\,mJy
\\
NGC 1068 $z=2$ \footnotemark[1] & $9.6\, \mu$Jy & $22\, \mu$Jy & $55\,
\mu$Jy & $0.16$\,mJy & $3.3$\,mJy & $16$\,mJy & $32$\,mJy \\ NGC 1068
$z=3$ \footnotemark[1] & $2.6\, \mu$Jy & $3.9\, \mu$Jy & $10\, \mu$Jy
& $32\, \mu$Jy & $0.86$\,mJy & $5.4$\,mJy & $12$\,mJy \\
M82 $z=2$ \footnotemark[1] & $74\, \mu$Jy & $0.11$\,mJy & $0.15$\,mJy &
$0.11$\,mJy & $1.4$\,mJy & $9.6$\,mJy & $58$\,mJy \\
M82 $z=4$ \footnotemark[1] & $4.2\, \mu$Jy & $10\, \mu$Jy & $19\,
\mu$Jy & $36\, \mu$Jy & $38\, \mu$Jy & $1.1$\,mJy & $5.3$\,mJy \\
HR10 $z=2$ \footnotemark[1] & $71\, \mu$Jy & $0.12$\,mJy & $0.15$\,mJy &
$0.11$\,mJy & $0.36$\,mJy & $6.6$\,mJy & $44$\,mJy \\
HR10 $z=4$ \footnotemark[1] & $2.4\, \mu$Jy & $6.6\, \mu$Jy & $19\,
\mu$Jy & $39\, \mu$Jy & $20\, \mu$Jy & $0.28$\,mJy & $5.0$\,mJy \\
\hline
\end{tabular}
\begin{tabular}{l}
\footnotemark[2]SWIRE limits from \citet{Lonsdale04}. \\ The SWIRE
team is currently re-assessing the observing strategy due to the
degraded MIPS sensitivities. \\ The values shown here for 70 and
160\,$\mu$m sensitivities are pre-flight estimates only.\\
\footnotemark[7]Confusion limit from observed source counts, of one
source per 40 SCUBA beams (7$\,^{\prime\prime}$ radius circle) \\
\footnotemark[3]Number of observed sources per SCUBA 850\,$\mu$m beam
(7$\,^{\prime\prime}$ radius circle) greater than the SWIRE limit. \\
\footnotemark[5]Number counts from \citet{Fazio}\\
\footnotemark[6]Number counts from \citet{Marleau}\\
\footnotemark[4]Number counts from \citet{Dole}\\
\end{tabular}
\end{table*}
\section{Data acquisition}\label{sec:method}
\subsection{The survey fields}\label{sec:fields}
The half square degree to be covered by SHADES is split between two
survey fields -- the Lockman Hole East (field centre approximately
$10^{\rm h}$ $52^{\rm m}$ $28^{\rm s}$ $+57^\circ$ $22^{\prime}$
$20^{\prime\prime}$) and the Subaru/XMM--{\it Newton\/} Deep Field
(SXDF, $02^{\rm h}$ $18^{\rm m}$ $00^{\rm s}$ $-05^\circ$
$00^{\prime}$ $00^{\prime\prime}$).  The fields were chosen for their
low Galactic cirrus (100\,$\mu$m surface brightness $\sim 1 {\rm MJy\,
sr^{-1}}$) and benefit from abundant cospatial multi-wavelength data.
The declination of the source fields and their spread in RA make these
fields observable for the majority of the year using the JCMT as well
as being accessible to BLAST, {\it Spitzer\/}, the VLA, UKIRT, Subaru,
Keck and Gemini telescopes.  The decision to include an equatorial
field was partly driven by the desire to provide a sub-mm source
catalogue accessible to ALMA.

Observations in the Lockman Hole East are being extended around the
151 arcmin$^2$ observed as part of the SCUBA 8-mJy Survey
\citep{Scott02}.  The direction and shape of this extension is driven
primarily by the existence of ultra-deep VLA imaging at 1.4\,GHz.
  
The specific choice of the SXDF as the equatorial field was motivated
by the existence of deep XMM--{\it Newton\/} imaging, associated deep
VLA observations at 1.4 \,GHz, and existing deep multi-colour optical
imaging obtained with the Subaru telescope.

The near-infrared ($J, H, K$) imaging for both fields will be provided
by the new UKIRT WFCAM instrument as part of the UK Infrared Deep Sky
Survey (UKIDSS).  Specifically, the Ultra Deep Survey (UDS) component
of UKIDSS will cover 0.77\, deg$^2$ to $K \simeq 23$ in the SXDF,
while the Deep Extragalactic Survey (DXS) in UKIDSS will map the
SHADES Lockman field to $K\simeq 21$.

Further multi-wavelength coverage comes from BLAST (500, 350 and
250\,$\mu$m) surveys in our fields and two {\it Spitzer\/} surveys --
the SWIRE Legacy Survey \citep{Lonsdale03, Lonsdale04} and the
Guaranteed Time Observations (GTO) data \citep{egami, huang}.  The
BLAST survey is expected to be confusion limited at all wavelengths,
corresponding to a $5\sigma$ flux density at the confusion limit of
approximately $25$\,mJy in all three bands \citep[derived
from][]{Rowan-Robinson}.  The SWIRE survey flux density limits are
shown in the first row of Table \ref{tab:sirtf}. Various
pointed followups are also underway including using SHARC-II at the
Caltech Sub-millimeter Observatory (CSO).

The BLAST 500\,$\mu$m data are expected to be deeper than the SCUBA
450\,$\mu$m data because the SCUBA observations are deliberately
restricted to be conducted in only grade 2--3 weather ($\tau_{\rm 225\,GHz}
\simeq 0.05-0.10$; this restriction was adopted to allow
other, smaller programmes on the JCMT to best exploit the rare, grade-1, 
conditions).  Because the 450\,$\mu$m SCUBA data will be of only
moderate quality, we expect few sources to be detected in the
450\,$\mu$m maps.  Therefore, in this paper, 
we focus primarily on the 850\,$\mu$m data.

\subsection{SCUBA technical information}
SCUBA is composed of two arrays of bolometers that view the same
region of sky simultaneously, a long wave array of 37 bolometers used
at 850\,$\mu$m and a short wave array of 91 detectors used at
450\,$\mu$m.  The pixels are arranged in a hexagonal pattern, with the
feedhorns close-packed.  The bolometer performance improves with
decreasing temperature, so SCUBA has a helium cooling system to
improve detector sensitivity.  Thermal noise from the sky and local
surroundings dominates at millimetre wavelengths.
 
The {\it SHADES} survey is conducted in jiggle-mapping mode.  The
SCUBA bolometers instantaneously undersample the sky but are dithered
in a 64-point pattern to ensure that, overall, the sky is sampled at
the Nyquist frequency or better at both wavelengths.  Further details
on SCUBA can be found in \citet{Holland}.

The terrestrial atmosphere and thermal emission from the telescope
both contribute to a strong background ($\sim$1\,Jy per square arcsec)
and the atmospheric part of this emission varies rapidly.  By rapidly
chopping the secondary mirror and nodding the entire telescope these
effects are reduced, leaving a residual atmospheric noise which is
common-mode to and therefore uniform across the whole array of
bolometers.

Jiggle-maps are coadded to improve the signal to noise ratio
(hereafter S/N) allowing measurements of signals that are tens to
hundreds of thousands of times fainter than the background.

\subsection{The observing strategy}
\begin{figure}
\centering { \includegraphics[width=8cm]{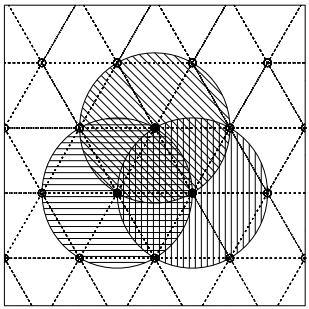}}
\caption { The `tripos' positioning scheme.  Individual jiggle-maps
($2.6'$ diameter) are represented as large hatched circles; only three
of these are plotted for clarity, but the centres of the others are
marked with small open circles.  This pattern of overlapping
jiggle-maps, in conjunction with observing different chop throw/PA
combinations at different airmasses, provides an approximately uniform
coverage over the map.  These three hatched jiggle-maps, with centres
marked by filled circles, comprise one `tripos'.  Only these three
positions contribute to the data within the triple-hatched triangle
formed by the filled circles. Any other triangular region is covered
by its own combination of three jiggle-maps.  The jiggle-map positions
shown by the open circles are observed sequentially in a spiral manner
from the centre of the map outwards; the resulting roughly circular
areal coverage minimises the perimeter, which has lower coverage and
so lower S/N than the rest of the survey.
\label{diag:tripos}}
\end{figure}

The survey makes use of close-packed hexagonal geometry to place
jiggle-maps in an interleaved positioning scheme.  This provides as
uniform a noise level across the survey field as practicable, as well
as ensuring that each sky position is covered by multiple bolometers.
Fig.\,\ref{diag:tripos} shows one interleaved `tripos' pattern,
illustrating the central triangular region covered uniformly by the
three hatched jiggle-maps.

The tripos system of three overlapping maps is observed for each of
six different chop-throw and chop position angle (PA) combinations.
This is a significant departure from the previous survey strategy most
similar to SHADES, that adopted for the 8-mJy survey \citep{Scott02,
Fox}.  We specifically chose the six chop throws motivated by the
Emerson-II chop throw methodology \citep{Emerson} i.e. chop throws of
30$\,^{\prime\prime}$, 44$\,^{\prime\prime}$, 68$\,^{\prime\prime}$,
at PAs of zero and 90 degrees in right ascension/declination
co-ordinates.  This choice of chop throw and PA ensures that no
Fourier modes larger than the beam and smaller than the largest chop
throw are lost entirely.  The source extraction from these multiple
chop throw maps is discussed below (Section \ref{sec:sourcextr}).  By
observing each of the chop throw/PA combination maps at different
airmasses, the noise levels in each tripos should ensure an even
coverage.  Thus, one chop throw / chop PA combination will always be
observed within a particular airmass range; while it would be possible
to balance the airmasses evenly over the chops in order to avoid
correlating the chop strategy with airmass, our adopted strategy makes
observing decisions easy enough to maintain a low error rate.  In
addition, the jiggle-maps at each tripos position are observed in a
priority order which keeps the survey area approximately circular at
any time, in order to minimise the perimeter area which will have
lower signal-to-noise.

In the SXDF, the observing strategy has been to extend the region from
the centre of the field in a spiral manner.  The area of Lockman Hole
East already covered by the 8-mJy survey has not been repeated, since
the noise level of this area is already at the required
depth.  Therefore, as discussed above, the new SCUBA data have been
extended around the existing 8-mJy Lockman Hole East data.

\section{The data reduction process}\label{sec:DataReduction}
We discuss here the Interactive Data Language (IDL) reduction pipeline
developed for SHADES from the original 8-mJy survey reduction process
\citep{Scott02, Serjeant03a}.  This process flat-fields the data,
combines the data from the individual chops and nods, and corrects for
the effect of the atmosphere using an extinction correction
routine.  The pipeline also corrects for noisy bolometers and fast
transient spikes in the data, e.g. cosmic rays (see below).  In order
to extract spatial information, the jiggle-map data can
be regridded onto rectangular co-ordinates.  The improvements and
variations on the 8-mJy survey pipeline made for the SHADES data are
listed below; for further details of this reduction method
\citep[see][]{Serjeant03a}.

\subsection{Extinction corrections}
Sky absorption caused by water vapour in the atmosphere was previously
removed from the data using smoothed CSO sky-dip measurements extrapolated to 
850 and 450\,$\mu$m (using the
conversion relations give by \citet{Archibald}) or by SCUBA sky-dip
measurements when insufficient CSO data were available.  CSO
tau sky monitors take sky-dip readings at 225\,GHz every 10 minutes.
SCUBA sky-dips have the benefit of being at the observation wavelength
but are taken at a different time to the observations.  The new
pipeline preferentially uses the water vapour radiometer data
\citep[WVM, ][]{Wiedner}, available since July 2000, that are taken at
the same time and azimuth as the observation at a frequency of 183\,GHz
every six seconds.  These WVM data have been converted into 225\,GHz
values, allowing for the elevation corrections, by Weferling
(priv. comm.).  The WVM data are smoothed over 18\, seconds -- a
timescale similar to that of the nod
-- and applied to the data.
We have undertaken a comparison of the WVM, SCUBA sky-dip and CSO sky-dip 
data, and find that they agree to within 10\% when
converted to 225\,GHz opacity.  This corresponds to derived flux-density
uncertainties of a few percent.  The sky extinction correction from the
WVM has the advantage of being able to reliably fit variations which
occur over the course of a single map as well as over the course of
the night.  It is therefore used for all nights where the necessary
data are available. 

\subsection{Bolometer and sky noise analysis}\label{sec:noise}
In what follows we refer to a single measurement (i.e. the average of
8 chop cycles subtracted between the 2 nod positions) from a single
bolometer as a `readout'.  Each readout represents approximately 2
seconds of integration time.

The noise in our data varies with time and between bolometers.  As the
noise was found to vary significantly on timescales significantly
longer that $\sim100$ readouts, we decided to measure the noise in
128-readout groups for each bolometer.  This is achieved by fitting a
Gaussian to the histogram of readouts of a 128-readout group.  Outliers
in these readout distributions at the $>3\sigma$ level are flagged as
glitches.  This filters out cosmic rays.  Instrument systematics should
be removed by the nodding technique.  We find, nevertheless, that
there is still residual sky emission in the data, caused by the sky
level varying on timescales less than that of the nod.  This level is
difficult to characterise, but appears to be common to all bolometers
in the array, taking into consideration the noise level of individual
bolometers.  The average in the form of the mode of the data is found
by fitting a Gaussian to the data from all bolometers for each of the
long/short wavelength bolometer groups for each time step.  The
DC-offset over the long/short array is then removed for each time
step.  Sky levels are evaluated independently at each wavelength so
that two flux readings can be taken at different wavelengths from the
data. Higher order terms of the sky-level are not removed in this
method as they have been found not to be significant.

\begin{figure*}
\centering{ \includegraphics[width=8cm]{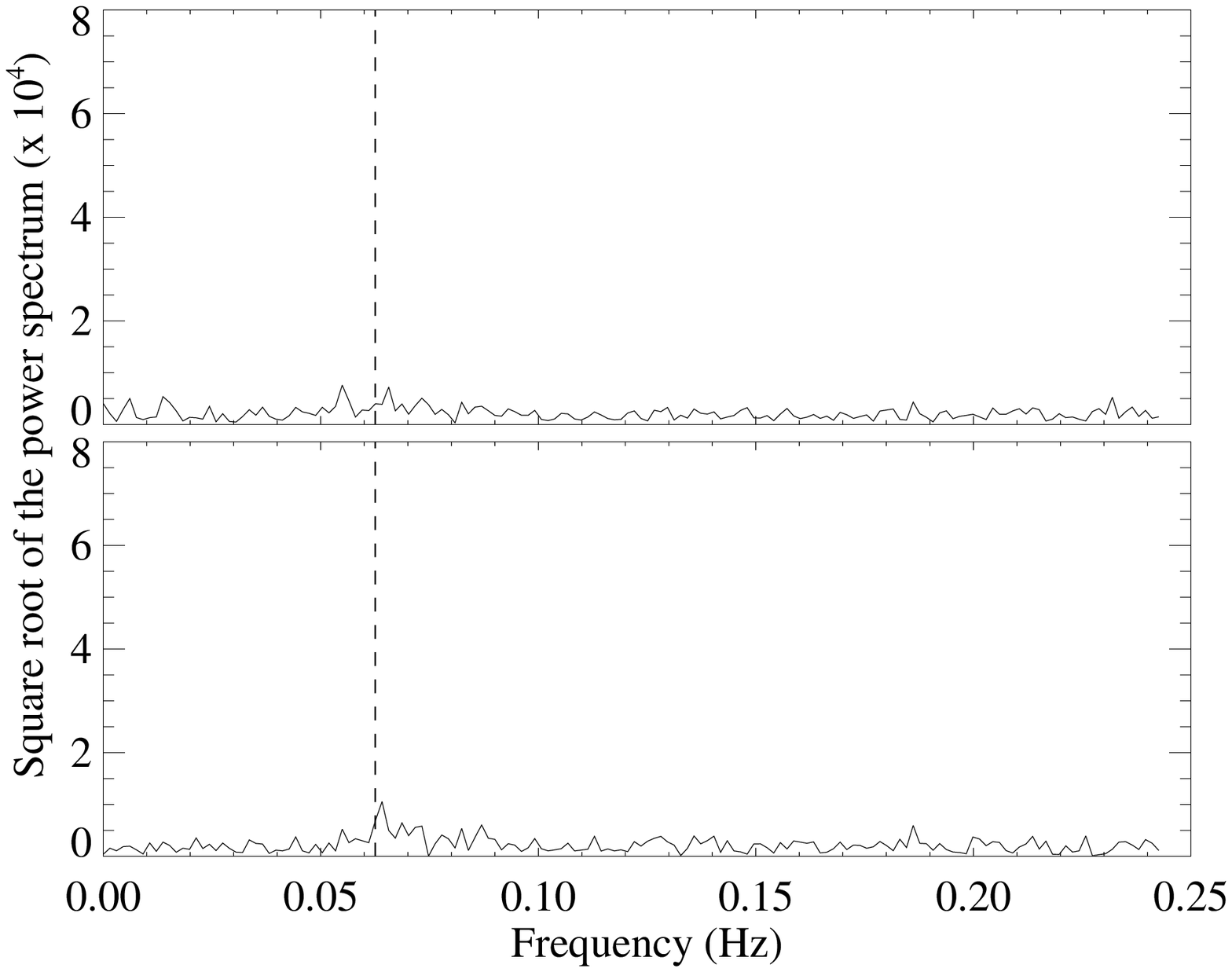}
\includegraphics[width=8cm]{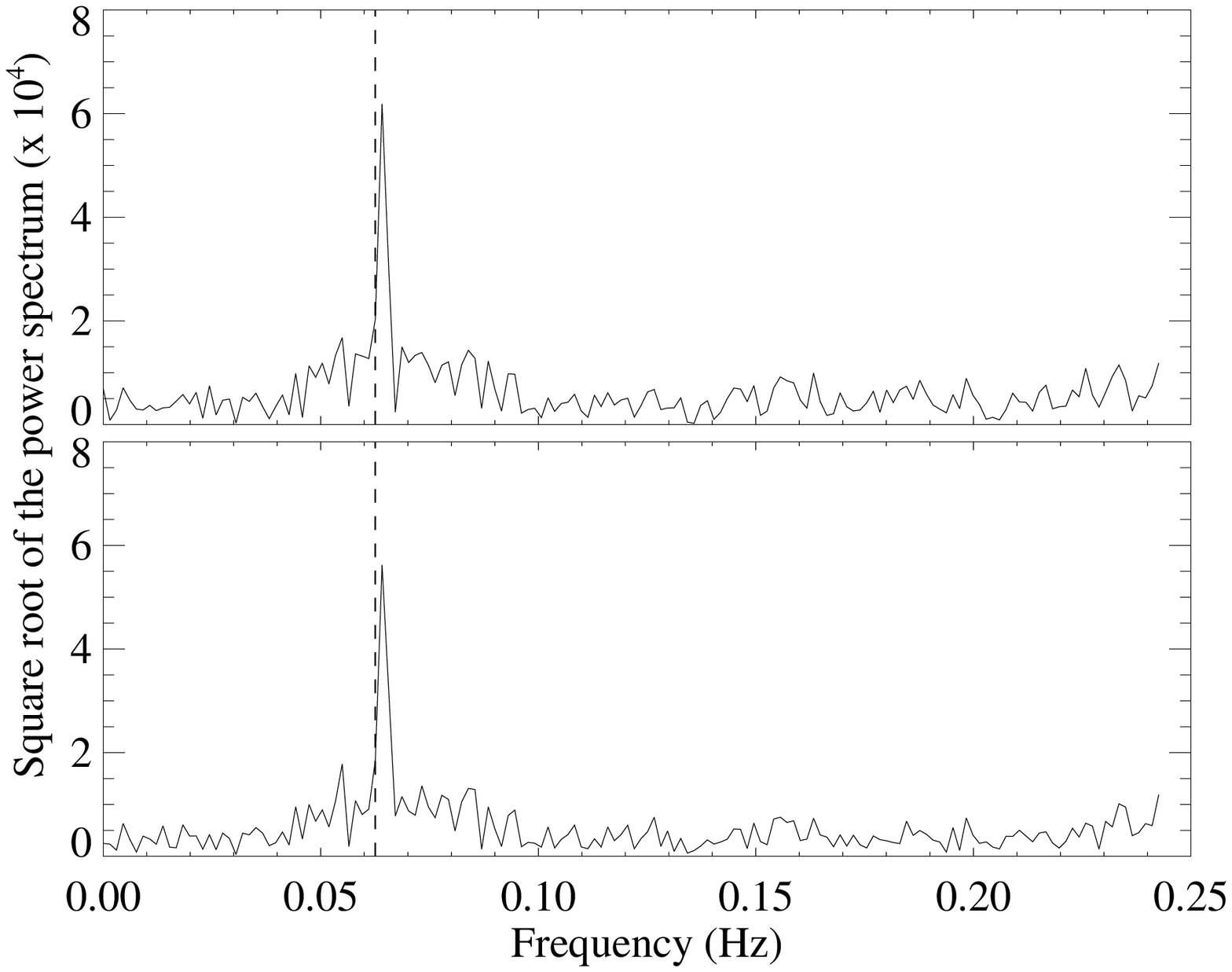} }
\vspace{1cm}
\caption{ The spike in the power spectrum of the bolometer readouts.
The upper row of plots shows the power spectrum of a typical (in this
case the central) bolometer, before (left) and after (right) sky
subtraction.  The lower row of plots shows the same for a bolometer
affected by the power-spike, showing clearly the spike at a period of
roughly 16 samples (at the dotted line) and the fact that the
power-spike itself has not changed.
\label{diag:powerspike}}
\end{figure*}

A significant difference from the 8-mJy survey is the presence of a
spike in the power spectrum of some of the bolometers at a period of
roughly 16 samples, for dates from the end of 2002 onwards as seen in
Fig.\,\ref{diag:powerspike} \citep{Borys04,Webb}. Tests have shown that
this effect is only significant for our data between December 2002 and
June 2003.  This effect presents difficulties for the sky-noise removal
program because the frequency is the same as that of the jiggle
pattern itself. It has the effect of skewing the distribution of the
sky-noise, so that it is not correct to remove the same DC sky level
for both those bolometers with the noise spike and those
without. Where a significant number (greater than 10) of bolometers
are affected in each map, the bolometers with and without the noise
spike are treated as individual data sets for the sky-subtraction
analysis. The process of bolometer noise measurement and sky
subtraction is then iterated.

In the method described in this paper we have not attempted
to remove the noise spike, but have simply chosen to isolate the affected
bolometers in our sky subtraction.  Therefore we do not knowingly
allow the presence of the spike to distort the sky subtraction
process.  As a result our assigned noise level 
s may not be optimal, although sources lost to the
higher noise regions should be accounted for statistically using
completeness simulations.  An alternative noise spike treatment, which
seeks to remove the noise spike from the data time-stream, 
has been applied to the SHADES data, and the results are almost identical.

Implicitly these algorithms assume the S/N of point sources to be
negligible in the bolometer readouts, which can readily be seen to be
the case given the thousands of readouts that contribute to the
3--10$\,\sigma$ sources in sub-mm surveys. However this is not true for
calibrators, so for these the noise measurement and deglitching stages
are replaced by assigning a noise value that consists of an arbitrary-
magnitude noise equivalent flux density (NEFD) value which scales as
the square root of the total integration time on each position on the
sky \citep{Archibald}.

\subsection{Flux calibration}
Calibration maps of Mars, Uranus, and secondary objects (the compact
non-variable sources 16293$-$2422, CRL2688, CRL618, HL Tau and the
variable sources IRC+10216 and OH231.8) have been used to calculate
the flux conversion factors (FCF) for each night
\citep{Jenness00}. The calibration maps are taken with the same chop
throws as the observations, and give information on the gain of the
telescope as it changes throughout the night in response to dish
temperature and environmental effects. The FCF is particularly
variable at times around sunrise/sunset.  Therefore calibration
observations at all three chop throws and positions are taken at the
start and end of each observing block, around sunrise/sunset.  Extra,
single chop throw, observations are also made when many hours of
observations have been taken without calibrating.

A source extraction routine identical to that used in the final source
extraction analysis is used to find the integrated voltage reading of
the calibration source, and this is compared to standard flux values,
including adjustments for the known variability of IRC+10216 and
OH231.8 \citep{Jenness02}, to calculate the FCF. This gives an FCF
representing ${\rm Jy\, V^{-1}}$ for the total flux of a point source
(see Section \ref{sec:srcextr}).  This calibration factor corresponds
to the typical notation of Jy per beam.  Calibration factors can have
an error of up to 10\%, because of variation with time and measurement
of the sources' integrated flux. The typical variation in FCF across
any SHADES shift is approximately 5\%, though this is larger when
data are taken around sunrise or sunset, or when observing is extended into 
daylight hours. The
FCF data are time-interpolated over the night, as opposed to using the
mean value for each half of the night, as was done for the 8-mJy
survey.

Some FCFs do not follow the trend of the data for the rest of the
night. The most common reason for this is because the calibrator is
extended. Mars behaved as a non-point source object due to its recent
proximity to the Earth and IRC+10216 has a CO/dust envelope of at
least 1 arcmin in extent.  These observations were not used to
calculate the night's FCFs.  Approximately 10\% of our calibration
maps are affected in this way.  When abnormal FCFs occur the
calibration changes throughout the night are tested using the pointing
observations of the secondary calibrators named above.

Where there are no usable calibration values or no agreement between
the pointing observations and calibration observations, standard
calibration values are used.  Standard calibration values were
evaluated using our own sample of calibrator and pointing maps,
reduced in IDL. These standard FCFs are approximately 20\% lower than
the JCMT standards using a SURF reduction, due to the differences in
the method used to rebin the data.  This systematic difference
demonstrates the importance of treating the calibrators in the same
way as the sources.  One test of these FCFs is to compare the flux
densities evaluated for sources with e.g. the flux densities
calculated using the 8-mJy survey-reduction method; the flux densities
were found not to be systematically affected.  Tests were also carried
out on the calibration observations to investigate the effect of a
noise-weighted source extraction as opposed to replacing the NEFD with
a fixed arbitrary value. The differences were much less than the
calibration factor errors for the respective wavelengths ($\sim 1\%$
at 850\,$\mu$m and $\sim 10\%$ at 450\,$\mu$m).

\subsection{Pointing corrections}
Telescope pointing observations are taken at intervals during the
evening to correct for the positioning of the telescope on a source
with respect to its pointing model.  Whilst these offsets are used
during data acquisition, drifts between pointing observations cannot
be corrected for at the telescope and so must be applied
retroactively. Once applied, the SCUBA2MEM program \citep{Jenness98},
takes them into account when exporting the positions of each bolometer
in the time series for further processing by the IDL pipeline.  Due to
the recent announcement of an error in the tracking model at the JCMT
which has affected data taken between August 2000 and April 2003
\citep{Tilanus}, we are making pointing corrections to the maps
currently reduced. The errors are in azimuth, but are elevation
dependent. $35\%$ of the maps covered in this paper were taken during
the affected period. However, the median absolute error of these when
combining the offset between a pointing observation and its science
observation is $0.53^{\prime\prime}$, with only $15\%$ of the affected
maps having an offset of $>1.5^{\prime\prime}$, which is of a similar
level to the usual rms pointing error of $1.3^{\prime\prime}$ in each
coordinate. The maximum error of any one map is 5$\,^{\prime\prime}$
and less than $0.5\%$ of our total maps have a tracking error of
$>3^{\prime\prime}$.

\subsection{Making zero-footprint maps}\label{sec:zerofootprint}

The final images are produced using an optimal noise-weighted
drizzling algorithm \citep{Fruchter} with a pixel size of 1 square
arcsec. This is the same method as that employed in the SCUBA 8-mJy
Survey \citep{Scott02} and Hubble Deep Field North \citep{Serjeant03a,
Borys} data reductions. Both output signal and noise maps were
created, the signal in any single pixel being given by the
noise-weighted average of the bolometer readouts for which this is the
closest pixel.  Unlike a standard shift-and-add technique, which takes
the flux density in each detector pixel and places it into the final
map over an area equivalent to one detector pixel projected on the
sky, drizzling takes the flux density and places it into a smaller
area in the final map. Although this significantly reduces the
signal-to-noise ratio in each pixel, this approach helps preserve
information on small angular scales, provided that there are enough
observations to fill in the resulting gaps. The area in the coadded
map receiving the flux from one detector pixel is termed the
footprint. Our method is an extreme example of drizzling in which the
footprint is selected to be as small as is practicable given the
pointing errors involved (termed the `zero-footprint'), effectively
placing delta-functions of flux into boxes one arcsec square.  This
also allows us to represent each bolometer's view of the sky as an
independent measurement, i.e. there is no pixel-to-pixel
crosstalk. The spatial variation in the sensitivity of the data is
taken into account during source extraction rather than at this
map-making stage. For the SHADES data, separate zero-footprint maps
are made for each chop throw and position angle.

A possible problem with this methodology is that it is lossy
compression: by coadding the data, individual pixels are combined
together and information about the distribution of flux readings is
lost.  However, it is possible during the running of the reduction
pipeline to calculate statistics that help evaluate the
self-consistency of the data contributing to each pixel and therefore
the reliability of the sources that are extracted from the map.

\section{The image processing procedure}\label{sec:sourcextr}

\subsection{Matched filter source-extraction techniques}\label{sec:srcextr}
Convolution of the image with a point spread function (PSF) is the
usual method for source extraction and is the optimal point source
filter in the case of uniform noise \citep{Heide, Eales99,
Eales00}. In the case of non-uniform noise, a method of minimisation
of $\chi^{2}$ of the data with the PSF is used, which can be expressed
as a convolution \citep[for further details see][]{Serjeant03a}.  This
method is optimal for point source sensitivity, but is not optimal for
spatial resolution and assumes that the sources will not be resolved
or confused. These assumptions can hold for the 450\,$\mu$m data, but
are more difficult for the 850\,$\mu$m data in which the beam size is
larger and confusion noise is more of a problem.

Indeed, both by examining the data and through 
clustering arguments it is clear that
some sources are partially blended and some are confused. For the
science goals of SHADES, it is important that sources fairly close
together on the sky can be properly separated, because otherwise much
of the potential power of the first radial bin in the angular
correlation function, $w(\theta)$, will be lost.

Starting from the simple case of unresolved point sources:
generalising the source extraction methodology of \citet{Serjeant03a},
the best fit (minimum $\chi^2$) point source flux at any point on the
sky is given by
\begin{equation}
F = { {\textstyle\sum_c} \; S_{c}W_{c} \otimes P_{c} \over
{\textstyle\sum_c} \; W_{c} \otimes P_{c}^{2} }
\end{equation}
where $F$ is the best fit flux, $c$ indicates the chop throw and
position angle combination, $S$ is the image signal, $W$ is the
reciprocal of image variance, $P$ is the point spread function, and
$\otimes$ indicates convolution. The S/N image is derived using
\begin{equation}
\frac{F}{\delta F} = { {\textstyle\sum_c} \; S_{c}W_{c} \otimes P_{c}
\over \sqrt{{\textstyle\sum_c} \; W_{c} \otimes P_{c}^{2}}},
\end{equation}
where $\delta F$ is the error on $F$. In our case, $P$ includes the
positive beam and both negative sidelobes.

Simulations shown in Figs.\,\ref{fig:sims1} and \,\ref{fig:sims2}
indicate that this multiple chop source extraction yields similar
S/N at the peaks as was inserted into the map at the source
positions, while removing the problem of negative-chop holes
coinciding with other sources.

\begin{figure}
\centering { \includegraphics[width=8cm]{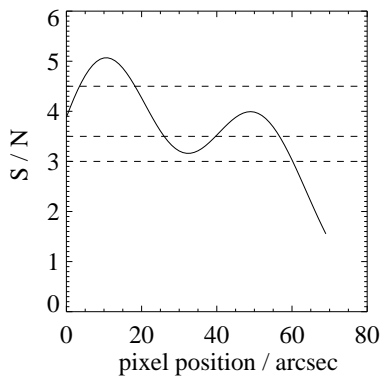}}
\caption {\label{fig:multicut} A connected pixels approach on
point-source-filtered data with multiple threshold cuts can identify
blended sources which have two separate peaks.  The 4.5$\sigma$ cut
would identify only the left hand peak, whereas the 3$\sigma$ cut
would only identify the right hand one.  The 3.5$\sigma$ cut would
identify two peaks.}
\end{figure}

Sources are currently identified as peaks in the S/N maps, using a
connected pixel approach using the IDL {\tt LABEL\_REGION} routine
(\url{http://www.rsinc.com}) to find the peak.
Sources were identified using multiple S/N threshold cuts between
3.0$\sigma$ and the peak S/N of the map. These multiple cuts allow
separation of blended sources, where there are separate peaks.  (see
Fig.\,\ref{fig:multicut}).

It can be shown that the method of multiple threshold cuts can
deblend Gaussian sources with equal fixed full width half maximum
(FWHM) having minimum separation between peaks of
$\sqrt{-2\,\sigma^{2} ln(p)}$ where FWHM $= 2\sqrt{2 ln 2} \sigma $
and $p = $ ratio of the individual sources' peak S/N where $0 < p\,\le
1$.  Sources are found with peaks $> 3.0\sigma$ and we have a maximum
S/N of $\sim$8.  We would therefore consider peaks with separation
less than $\sim$8.33$\,^{\prime\prime}$.  This is comparable to the
beam size of the SCUBA beam and therefore this should only become a
significant issue if sources are found with much higher S/N. For the
purposes of our completeness and reliability simulations, and for our
future source count derivations from this analysis, we only use the
method of multiple cuts described above.  

Since source counts are not the only science goal, it may
be useful to attempt deblending of source pairs
which are closer than the beam width.  To that end a more
sophisticated source deblending algorithm has been developed, building
on that detailed in \citet{Scott02}.  In the Scott et al. technique
the flux densities of all significant peaks are fitted simultaneously
using a maximum-likelihood technique at the positions of the peaks. We
have extended this to fit to the spatial position as well as the flux,
using the position and fluxes of sources from the source-extraction
method as a starting point and fitting between one and six sources at
each position. This becomes a computationally tractable problem when
we consider only the pixels around the $> 3\sigma$ peaks in the map
that would contribute to the flux at that position.

Adding additional sources to the fits can improve the reduced
$\chi^2$, but these new sources may not necessarily represent the
underlying distribution. Distinguishing between two close point
sources and some other extended structure (such as a single extended
source) implicitly involves constraining the quadropole moment of the
image, and as \cite{Lucy} has shown, the signal-to-noise requirements
for constraining the higher order multipoles are extremely stringent.
Nevertheless, the positions of known sources may be refined using our
multi-wavelength data. Such refinement is optimal if one
simultaneously fits all the relevant multi-wavelength data, and this
can be done within our methodology. 

As an example of this methodology, we have attempted to
deconvolve Lockman 850.1 and 850.8 using the 850\,$\mu$m maps from the
8mJy survey. The reported fluxes of these sources in the 8\,mJy survey
are $10.5\pm1.6$ mJy and $5.1\pm1.3$\,mJy respectively . We performed
the multiple source fitting, and the best fit is given in table
\ref{tab:lock1and8}, the covariance matrix of which shows a weak
correlation of fluxes with positions and between the fluxes of the two
sources.

\begin{table}
\caption{\label{tab:lock1and8} \textbf{The best-fit parameters for
fitting two sources to the area around Lockman
850.1 and 850.8. The position and flux errors are formal 1\,$\sigma$
errors computed from the covariance matrix.}}
\begin{tabular}{lll}
Position & Flux (mJy) & S/N \\ \hline 10 52 01.31 +57 24 44.4 &
5.2$\pm$1.1 & 4.9\\ 10 52 00.33 +57 24 19.0 & 9.3$\pm$1.2 & 8.0 \\
\end{tabular}
\end{table}

\subsection{Emerson-II deconvolution}\label{sec:emerson}

An alternative method to construct images is to use the multiple chop
strategy to recover the modes missing in any single chop, through the
Emerson\,II deconvolution algorithm \citep{Emerson, Jenness00}.  This
has been shown to be effective in reproducing sources in data with two
chop throws and one position angle, as was the observing strategy
applied to the Hubble Deep Field \citep{Hughes98, Serjeant03a}.  This
is not the only possible methodology, but it has the advantage of
having clear precedents in Galactic astronomy
\citep[e.g.][]{Johnstone, Pierce-Price}.

\begin{figure*}

\centering { \includegraphics[width=8cm]{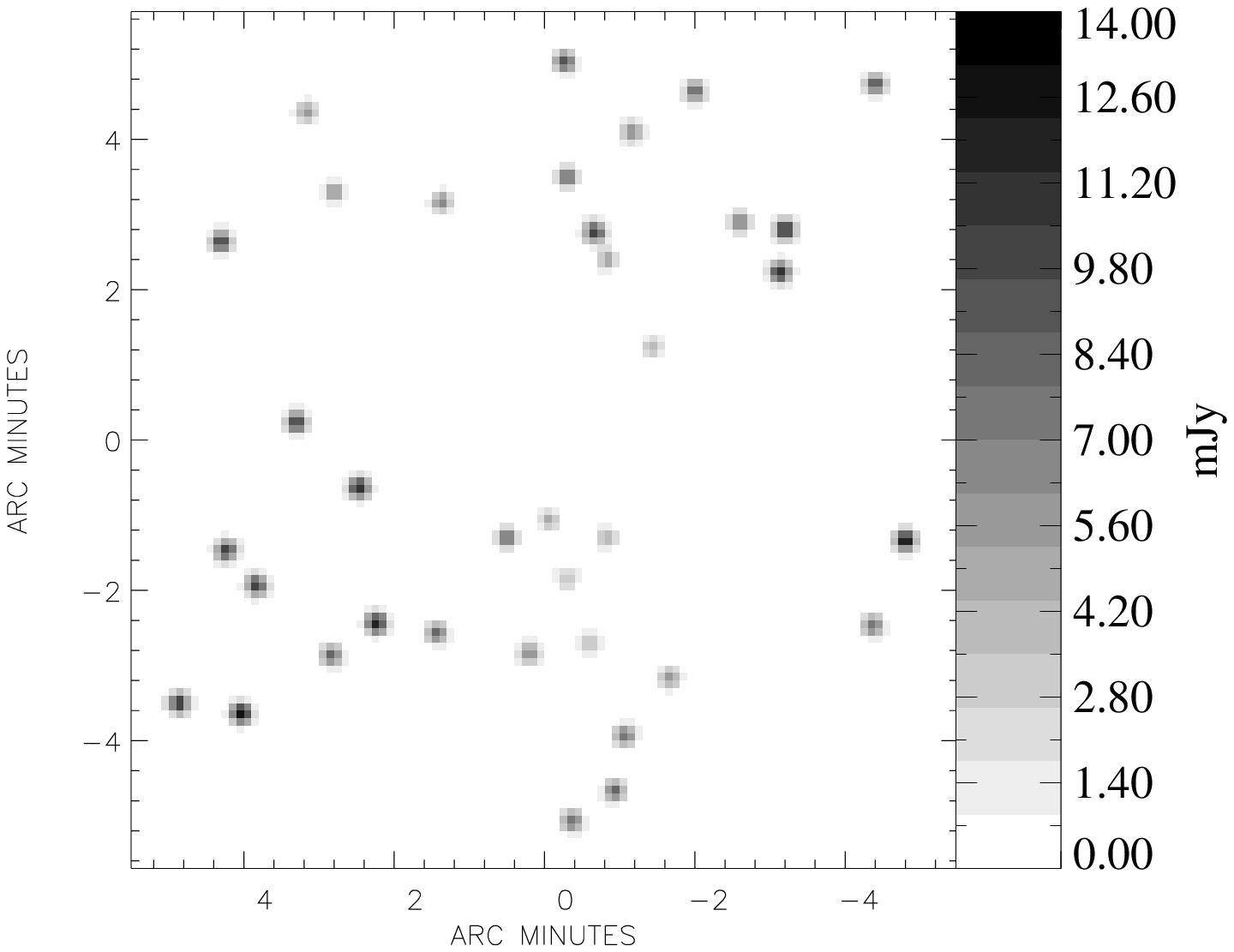}
\includegraphics[width=8cm]{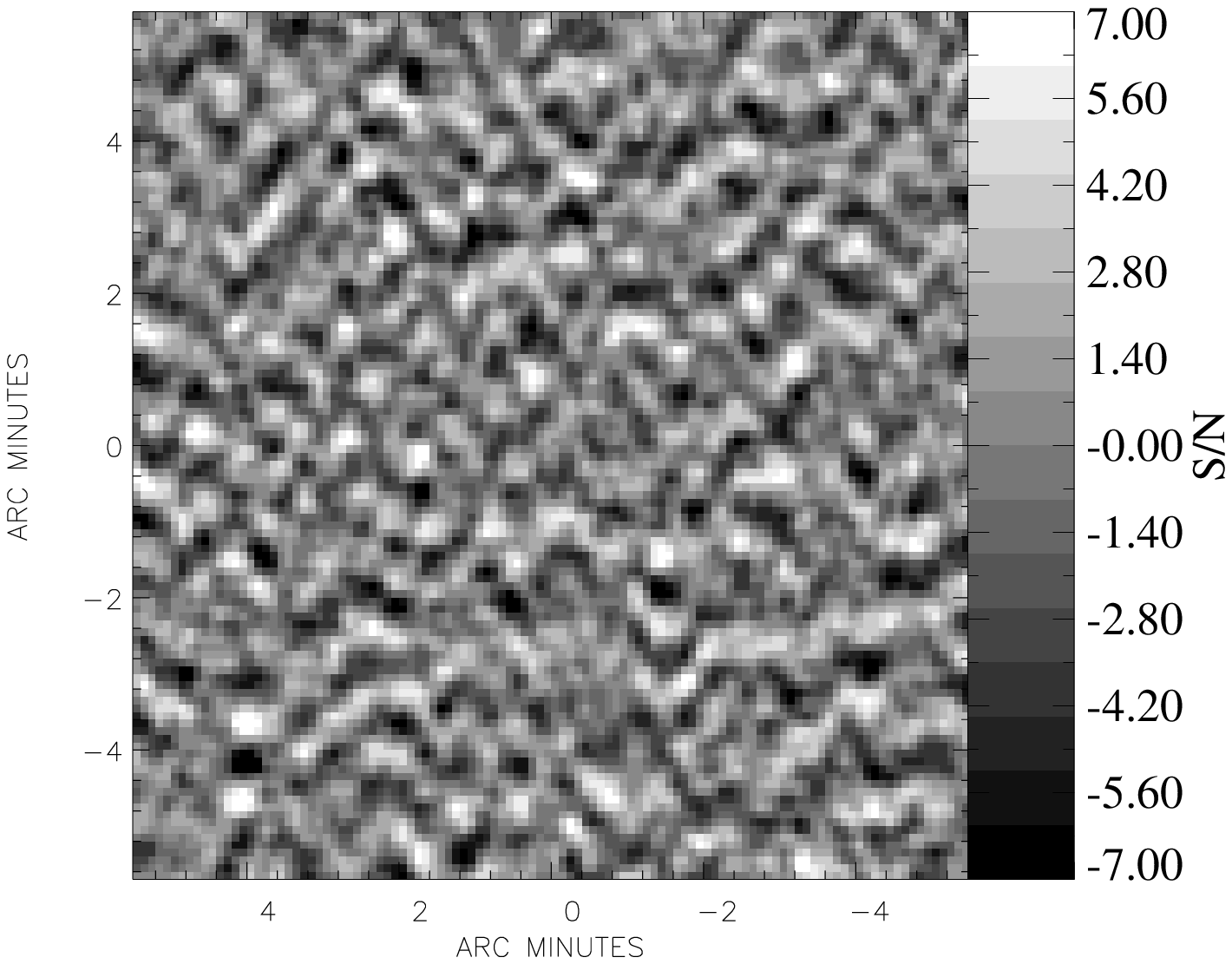} \includegraphics[width=8cm]{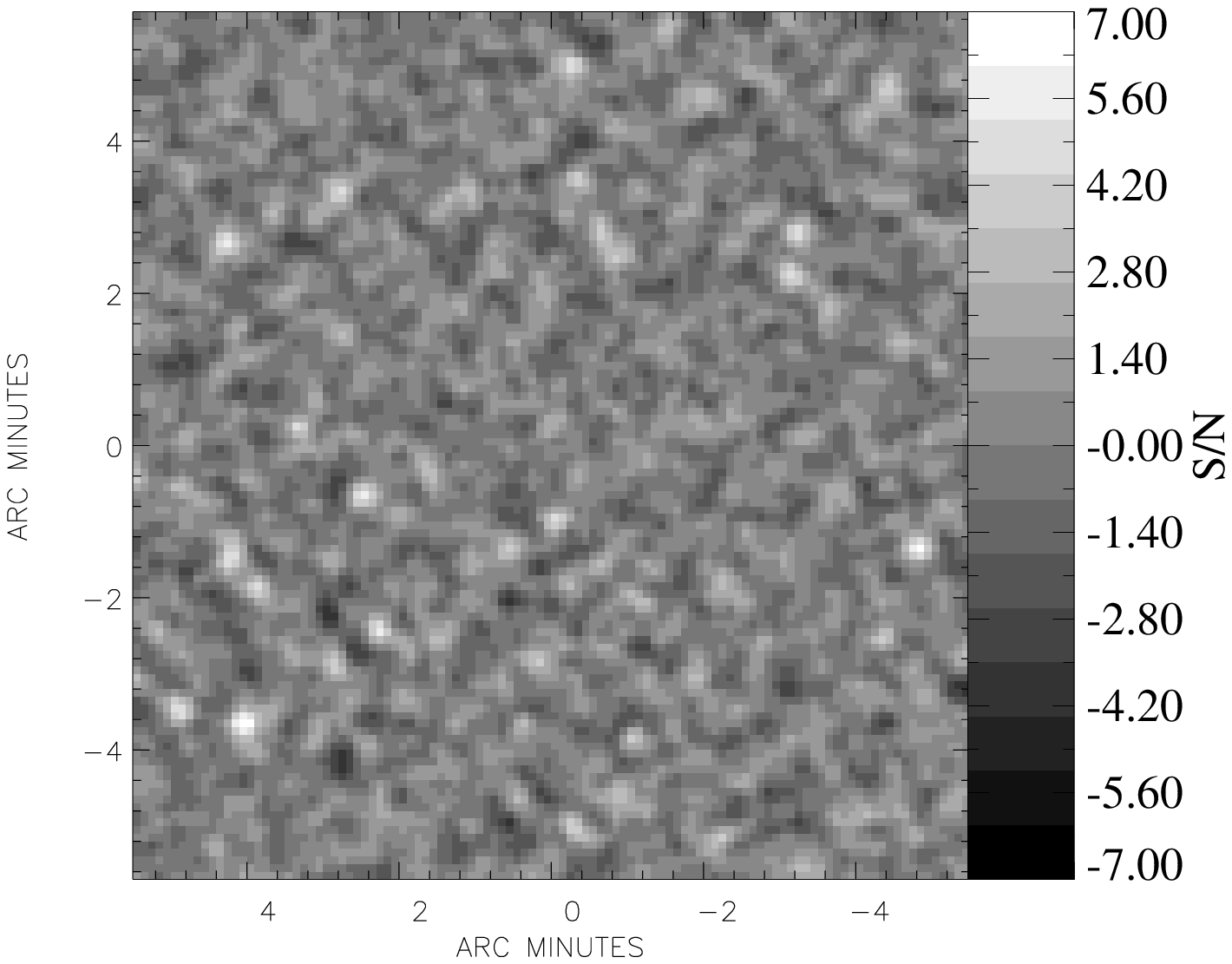}
\includegraphics[width=8cm]{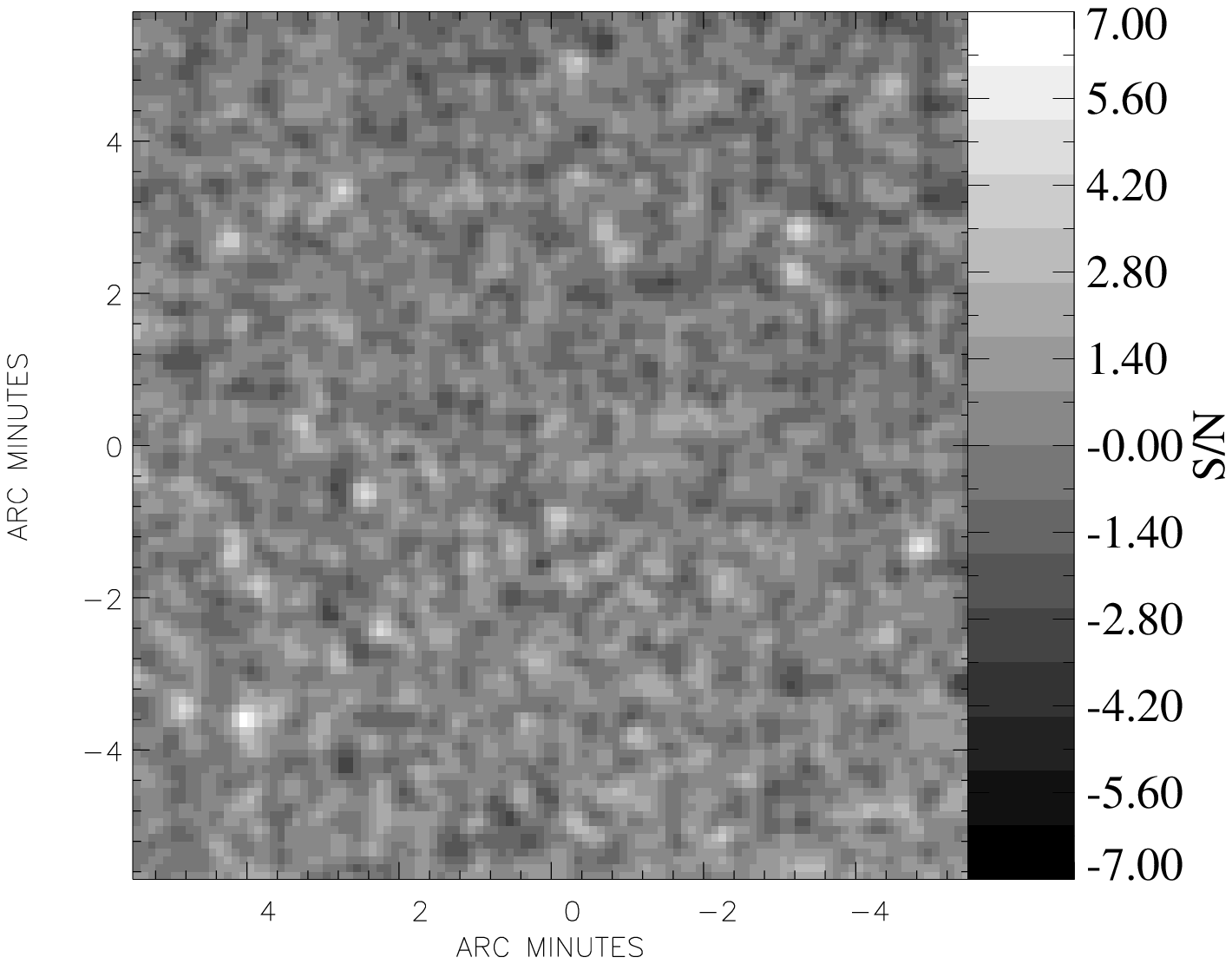} }
\caption { A simulation of source-extraction using the multi-chop
technique.  {\it Top left:} The input sources as they would appear
with no noise or chopping.  {\it Top right:} A source extracted map
using only one chop throw/PA image.  This is the method that was used for
the 8-mJy survey, such that the source extraction routine causes a
double chop function to be seen. This causes blending of some of the
sources with the chop functions of other sources.  {\it Bottom left:}
A source extracted map using six chop throw and PA combinations, each
with the individual map having $\sqrt{6}$ times the noise used in one
chop image.  {\it Bottom right:} The same simulated map was used in an
Emerson-II deconvolution method to produce a map without significant
chop residues.
\label{fig:sims1}}
\end{figure*}

\begin{figure*}

\centering { \includegraphics[width=8cm]{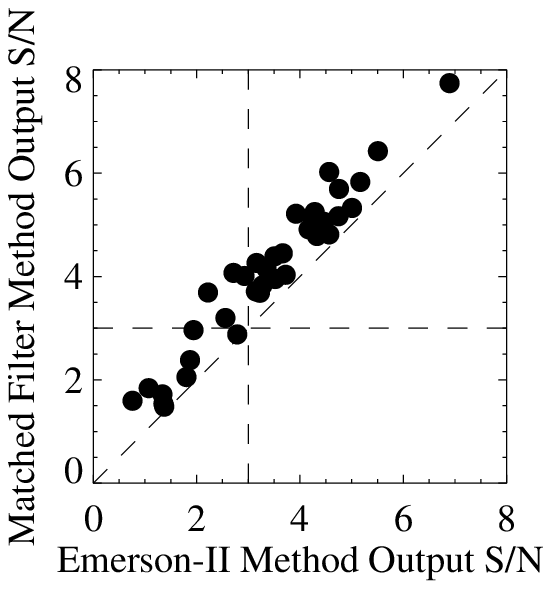}
\includegraphics[width=8cm]{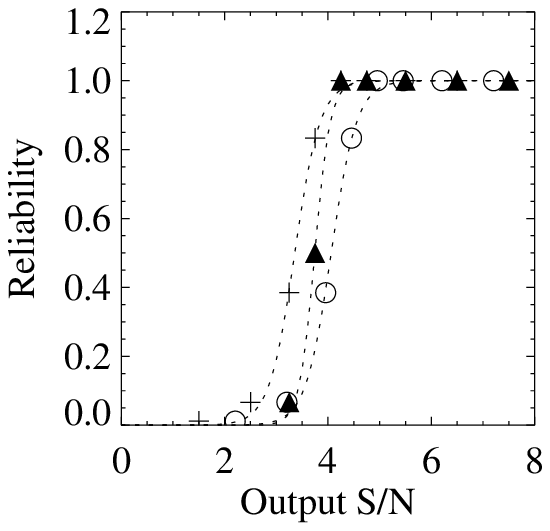}}
\caption { {\it Left:} Comparison of the output S/N of
the input sources for the matched filter and Emerson-II deconvolution
methods, for the simulations shown in Fig. \ref{fig:sims1}.  The
dashed lines show S/N = 3 for each of the methods, and the line for
which the two methods would have equal S/N.  {\it Right:} The
reliability of sources extracted from the matched-filter method
(filled triangles) and Emerson-II deconvolved maps (crosses). This is
based only on the sources in the top-left panel, which is far fewer
than the more extensive simulations in Fig. \ref{fig:simres};
however, since it is the same sources being extracted in both
matched-filter and Emerson-II methods, the {\it relative}
reliabilities can still be usefully compared. For a fixed S/N the
reliability appears to be higher for the raw Emerson-II deconvolution,
but this is easily shown to be an artefact of the poorer S/N of
Emerson-II for a fixed input flux (left).  After adjusting for the
poorer S/N in the Emerson-II method (open circles), the reliability is
shown to be less for the Emerson-II deconvolved maps than the 
matched-filter method. Thus, for example, a
3.5\,$\sigma$ source in an Emerson-II map is intrinsically brighter
than a 5\,$\sigma$ source in a matched filter map, and it is
unremarkable that
such brighter sources can be extracted more reliably; however, an
8\,mJy source extracted from an Emerson-II map has a lower reliability
than an 8\,mJy source from a matched-filter extraction.
\label{fig:sims2}}
\end{figure*}

The difference image produced by the chop process during observation
can be considered a convolution of the sky with the chop
function. Recognising that a convolution in real space is the same as
multiplication in Fourier space, the Emerson-II algorithm reconstructs
the image by effectively dividing by the chop function in Fourier
space.  Using this technique, modes in Fourier space are lost where
the Fourier transform of the chop function is zero, but these can be
filled in using modes from data taken at other chop angles.  This
method has the benefit of using the flux from the negative chop-holes
and {\it folding\/} it back onto the source position.  The down-side
of this for the Hubble Deep Field data is that modes are lost
altogether because the $30^{\prime\prime}$ and $45^{\prime\prime}$ are
in a 2:3 ratio.  In contrast, the chop strategy implemented for SHADES
is well suited to this reduction method because the chop throws are
incommensurate and the particular values of $30^{\prime\prime}$,
$44^{\prime\prime}$ and $68^{\prime\prime}$ have been shown previously
to work well with the Emerson-II technique.  Fig.\,\ref{fig:sims1}
shows an example of the source-extraction method of the primary
pipeline, in which a number of sources, simulated using the 8-mJy
survey source positions and fluxes, have been extracted using the
multiple-chop source extraction described above.  Fig.\,\ref{fig:sims1}
also shows the alternative method of reconstructing the image using
the Emerson-II algorithm \citep[the analogous algorithm for SCUBA scan
maps is discussed in][]{Johnstone}.These two algorithms give
superficially extremely similar reconstructions; this is partly
because the multiple chop throws and position angles distribute the
negative sidelobe fluxes over a large number of positions.  Although
the Emerson-II method may be useful for producing cosmetically clean
images, we have not investigated the presence of possible artefacts
made by the map-making process, and we prefer to use the direct method
outlined in Section \ref{sec:srcextr} to find sources.

Another approach is to use an iterative reconstruction scheme,
motivated by Cosmic Microwave Background map-making methods
\citep[e.g.][]{Wright96} and successfully applied to SCUBA scan-maps
by \citet{Johnstone}. The triple-beam pattern used in jiggle-map mode
is more problematic for this approach, since each datum is the
difference between one map value and the average of two others.
Investigation of the iterative reconstruction of SHADES maps has been
of only limited success \citep{Lepage04}.


\subsection{Comparison of these two methods} \label{sec:compsrcext}

To assess the relative merits of these two methods we compare the 
completeness and reliability of sources found using simulations (see
Fig.\,\ref{fig:sims2}).  The simulations have two components. 
First, a single
Gaussian chop profile source is added at a random position in the real
zero-footprint chop maps.  We then attempt to recover the source
(within $7^{\prime\prime}$ of the input position and within a factor
of two of flux and S/N $>$3.5) and information about the position and
flux of the output source is retained.  This gives information about
the completeness and flux boosting.  In the second method, 
a map is entirely simulated using the source counts 
of Scott et al. (2005), the level of clustering seen in the 8-mJy 
survey (Scott et al. 2002), 
and Gaussian random noise is added using the real noise
map as input. Sources are then extracted as described above and the results
used for calculating the reliability of the extracted sources, which
we here define as the number of 'real' sources found as a fraction of
the total number of sources found by the source identification
process.

The reliability of the Emerson-II method appears higher for a given S/N
(crosses in Fig.\,\ref{fig:sims2}) than the matched-filter method
(filled triangles), but this is because the S/N for an individual source
is lower in the Emerson-II method. After correcting for this
S/N difference (open circles) it is clear that the matched filter has
a higher reliability than Emerson-II for a given source (i.e., rather
than a given S/N). 

\section{Results and analysis}\label{sec:analysis}
\subsection{Progress}
A total of 1843 individual jiggle-maps, observed over 139 nights up to
February 2004, have been coadded into zero-footprint maps covering 720
arcmin$^2$.  The median noise level in the 850\,$\mu$m data is 2.2\,mJy
and that of the 450\,$\mu$m data is 25\,mJy. The data were taken within
a range of weather conditions (JCMT grades 2--3) such that the mean
$\tau_{850}=0.27\pm0.07$ and $\tau_{450}=1.45\pm0.44$. 92\% of the
data have $\tau_{850}$ in the range
0.20--0.48
and $\tau_{450}$ in the range 0.94--2.78.

\begin{figure*}
\centering { \includegraphics[width=8cm]{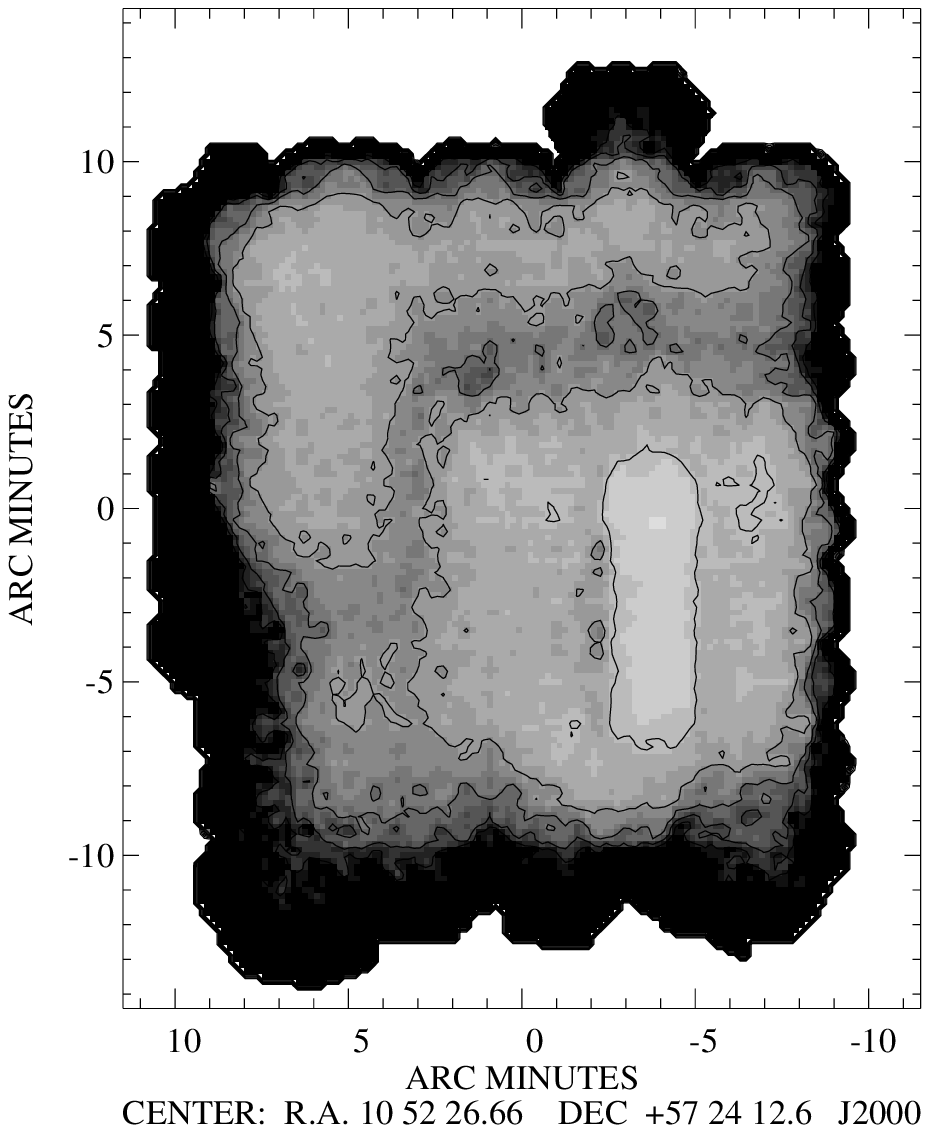}
\includegraphics[width=8cm]{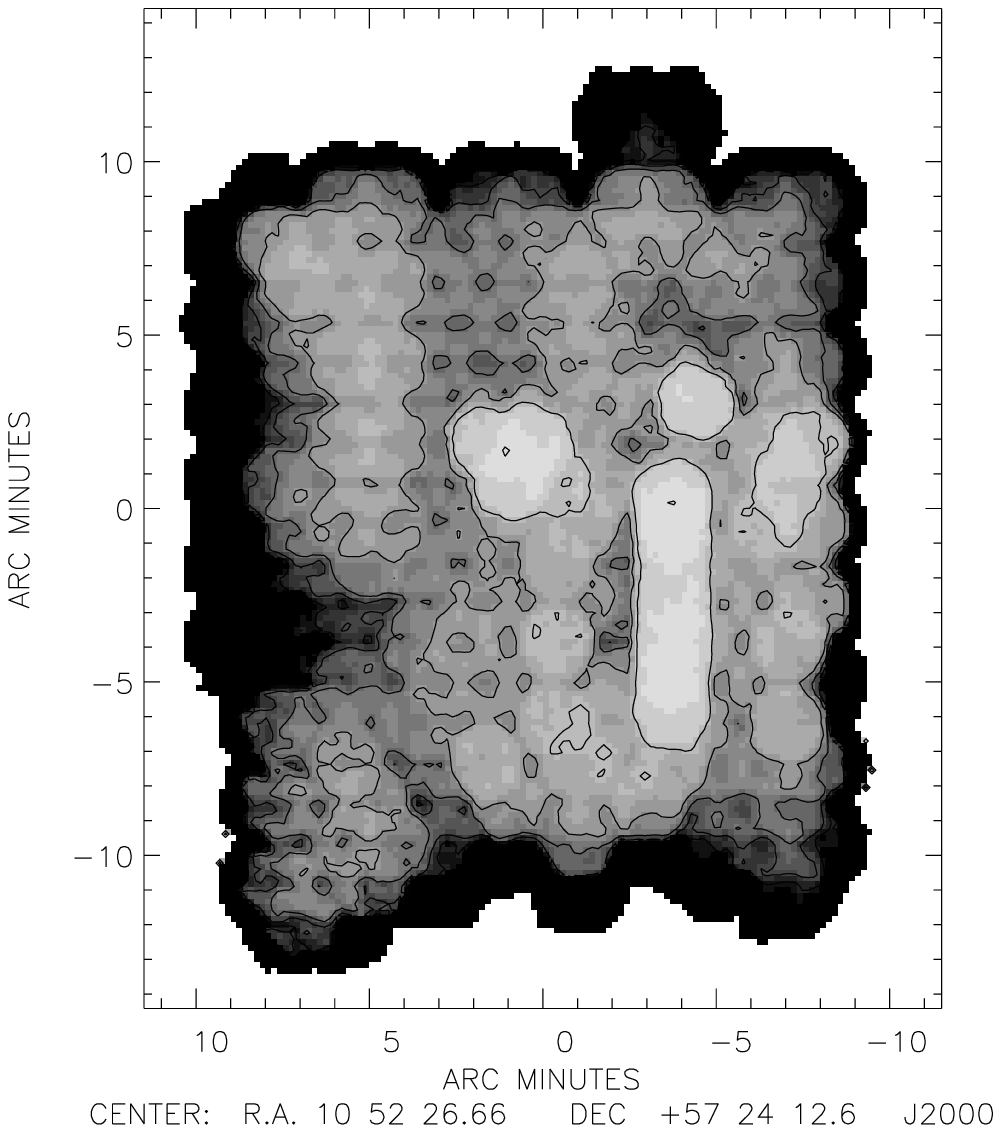}
\includegraphics[width=8cm]{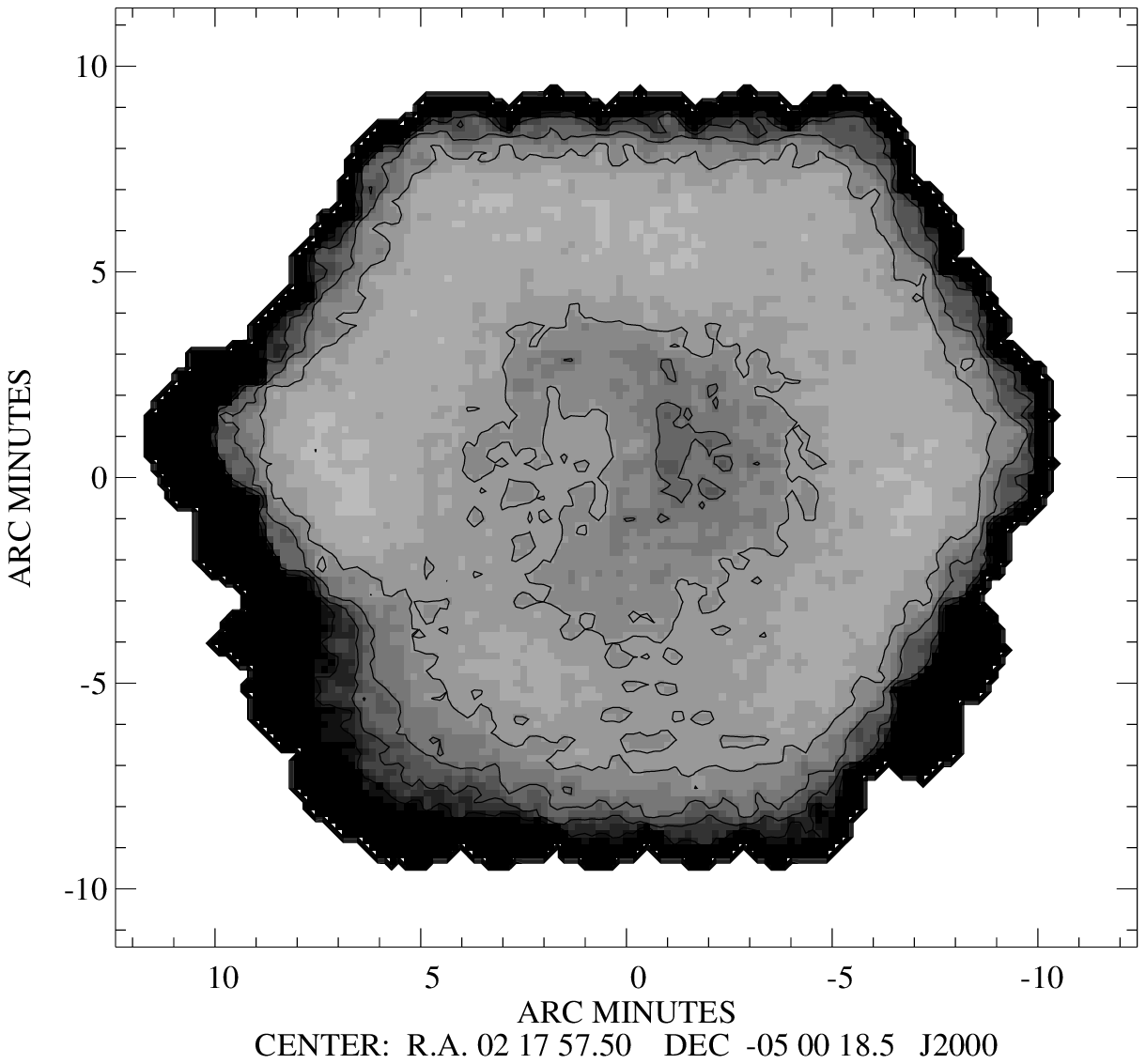}
\includegraphics[width=8cm]{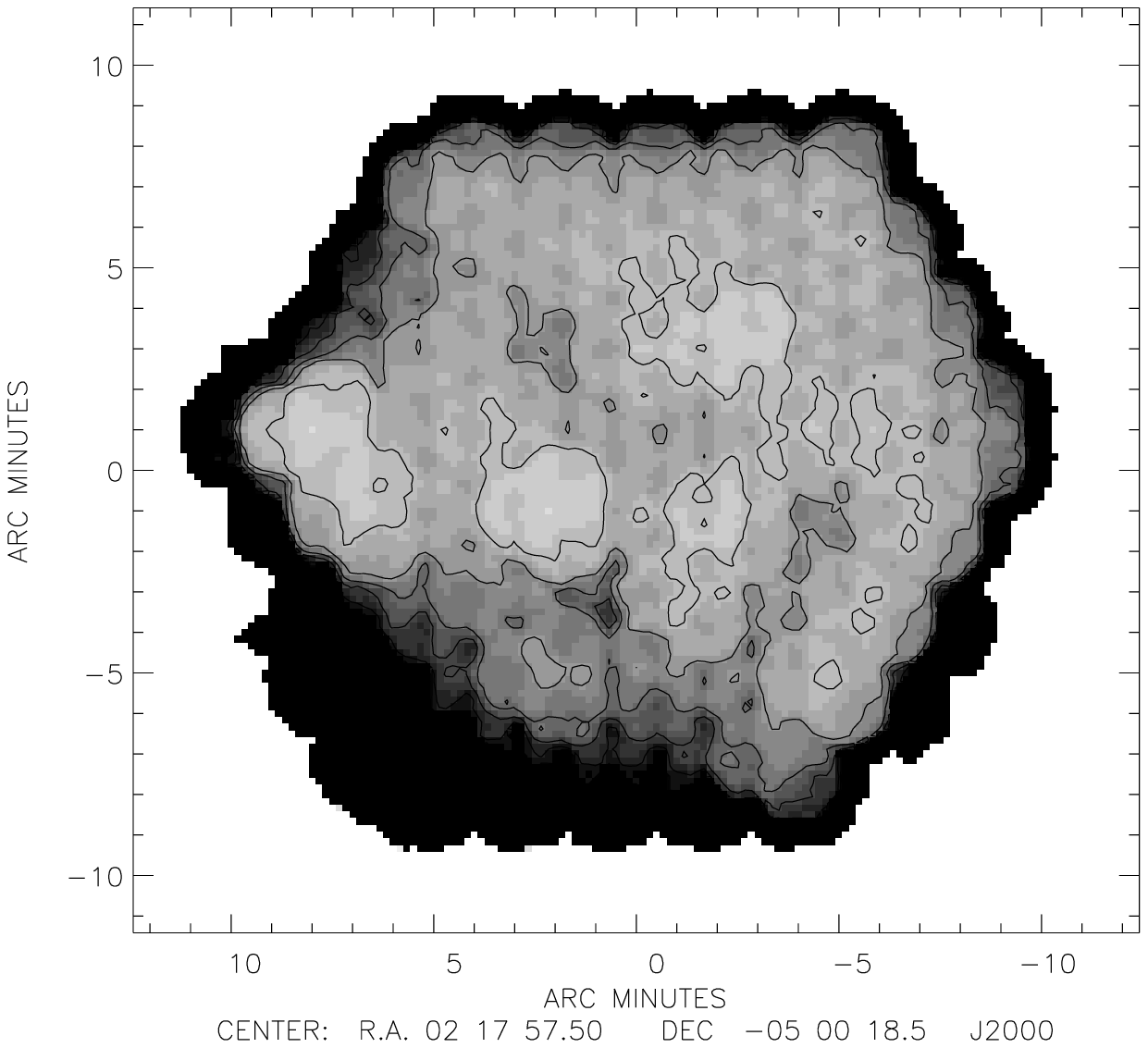}}
\caption { Noise maps of the ({\it Top Row\/}) Lockman Hole East and
({\it Bottom Row\/}) SXDF for data taken up to February 2004.  For the
850\,$\mu$m data ({\it Left Column\/}), the lowest contour level is
0.71\,mJy with higher
contours spaced by 0.71\,mJy to a maximum level shown in the map of
5\,mJy.  For the 450\,$\mu$m data ({\it Right Column\/}), the lowest
contour level is
7.1\,mJy with higher
contours spaced by 7.1\,mJy to a maximum level in the map of 50\,mJy.
Note the high levels of noise around the edges of the map, which will
be reduced when more observations are taken. 
\label{fig:noise_plots_all}}
\end{figure*}

The areas covered by the survey to February 2004 are shown in
Fig.\,\ref{fig:noise_plots_all}.  Note the uniformity of the noise
levels within the central regions of the maps as designed, with the
higher noise regions visible around the perimeter where further
observations are due to be taken.  Although the SHADES data are taken
only within the fixed weather conditions outlined above, the data
taken at the start for the SXDF were taken with consistently poorer
than average weather (grade 3), hence the higher noise level at the
centre of the SXDF map. However, the depth is consistent across that
area and varies only by $\simeq 0.7$\,mJy from the rest of the map.
Also note the deep strip in the Lockman Hole East, taken during the
SCUBA 8-mJy survey.

The source-extraction method can also yield sources near
the edge of the map which we choose to reject due to insufficient
coverage/sampling and/or high noise values. These can be identified
and removed from the source lists by creating a mask of the
integration time convolved with the beam (which in our case rejects
those regions near the edge of the map with little coverage) and
rejection of those sources with large flux-density 
errors (typically greater than
10\,mJy at 850\,$\mu$m and greater than 100\,mJy at 450\,$\mu$m).

\begin{figure*}
\centering { \mbox{\includegraphics[width=8cm]{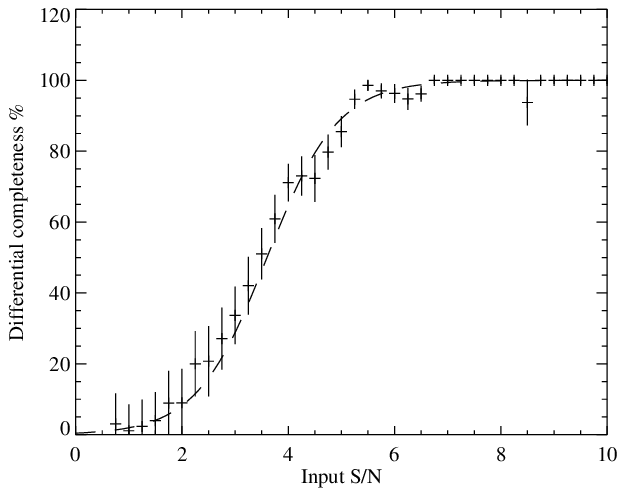}}
\vspace{-3cm} \mbox{\includegraphics[width=8cm]{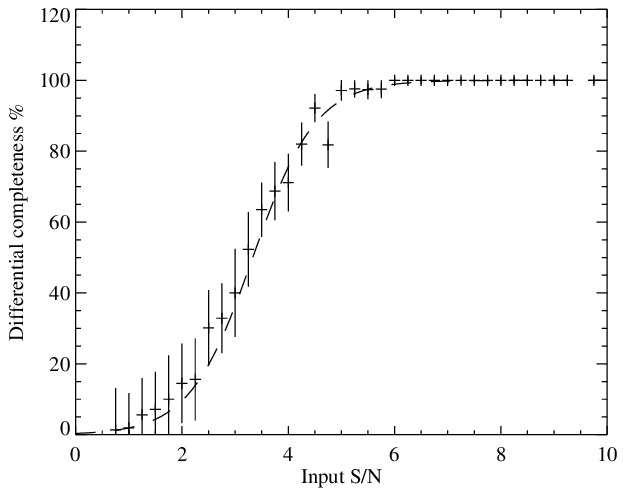}}
\vspace{-3cm} \mbox{\includegraphics[width=8cm]{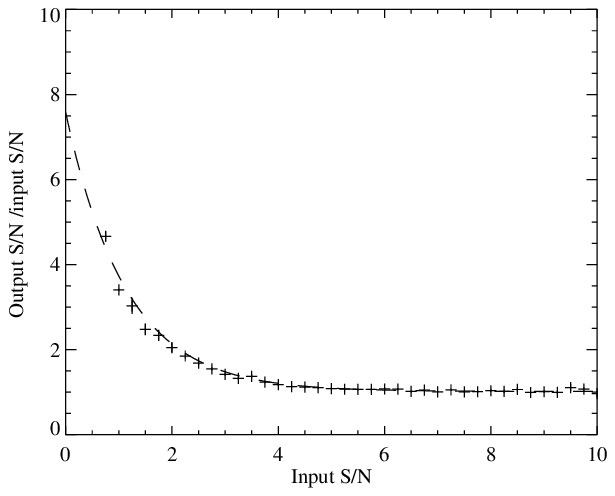}}
\mbox{\includegraphics[width=8cm]{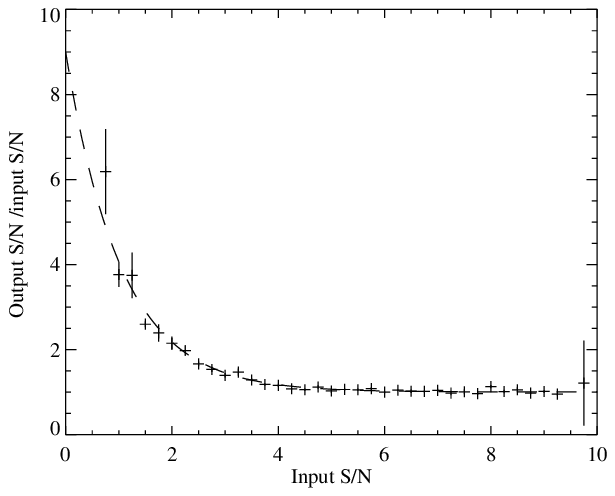}}
\mbox{\includegraphics[width=8cm]{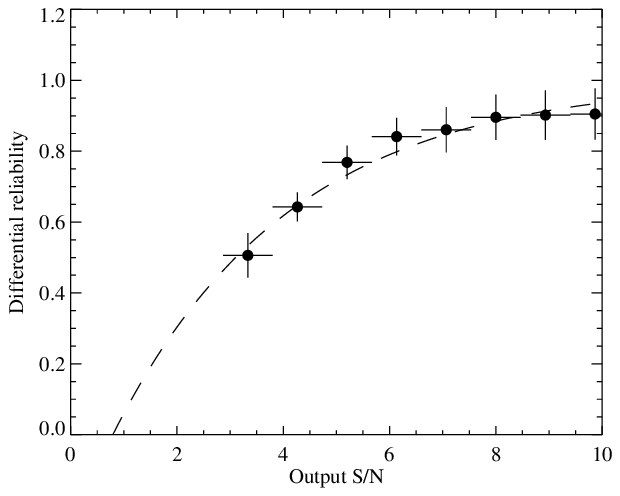}}
\mbox{\includegraphics[width=8cm]{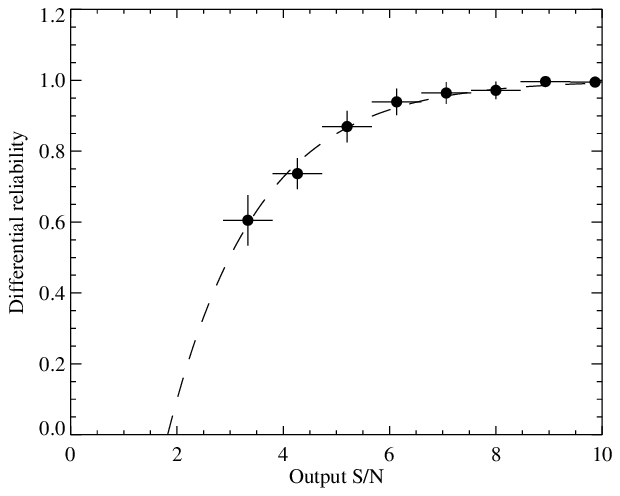}}
\caption { Completeness (top), flux boosting (middle) and reliability
(bottom) for the Lockman Hole East (left) and SXDF (right). }
\label{fig:simres}}
\end{figure*}

Using these criteria, simulations were carried out as described above
in section \ref{sec:compsrcext} for the full SHADES interim maps.
These gave completeness, reliability and flux boosting
effects as shown in Fig. \ref{fig:simres}.  The expected completeness
and reliability at 3.0, 3.5, and 4\,$\sigma$ are shown in Table
\ref{tab:reliability}. 

\begin{table}
\begin{tabular}{lllll}
\hline & $R$ & $C$ & $B$ & $D$ ($\,^{\prime\prime}$) \\ \hline LH & &
& & \\ \hline $3.0\sigma$ & 48 \% & 29\% & 1.47 & 2.90 \\ $3.5\sigma$
& 56 \% & 46\% & 1.31 & 2.85 \\ $4.0\sigma$ & 62 \% & 65\% & 1.21 &
2.79 \\ \hline SXDF & & & & \\ \hline $3.0\sigma$ & 50 \% & 36\% &
1.45 & 2.89 \\ $3.5\sigma$ & 63 \% & 57\% & 1.28 & 2.73 \\ $4.0\sigma$
& 73 \% & 76\% & 1.18 & 2.57 \\
\end{tabular}

\caption{\label{tab:reliability} Comparison values of reliability
($R$), completeness ($C$) flux boosting factor ($B$) and positional errors
($D$) at different threshold cuts for our two fields. }
\end{table}

Table \ref{tab:numsnr} gives the number of sources at
different significance cuts for each of the SHADES fields using the
source-extraction method outlined above.  In order to estimate the
source-detection density, we have conservatively considered only those
sources that are also found in two independent reductions, our
so-called consensus list (which will be discussed in future papers).
This in effect means that, for sources of modest significance in the
range 3--3.5\,$\sigma$, we only consider those that are reproduced in
all four reductions undertaken so far.  

In the Lockman Hole East map, 69 sources have been identified with
${\rm S/N}\geq 3$ of which 47 have ${\rm S/N}\geq 3.5$.  In the SXDF,
61 sources have been identified with ${\rm S/N}\geq 3$, of which 53
have ${\rm S/N}\geq
3.5$.
Thus, with 40\% of the data taken, SHADES has
produced a sample of 100 sources at 850\,$\mu$m with ${\rm S/N}\geq
3.5$.

The implied surface density of sources with ${\rm S/N}\geq 3\sigma$ is
therefore $653\pm57$ sources deg$^{-2}$.  The error on the estimated
source density is calculated using the Poisson error on the data
(i.e. the number of sources observed), corrected for the area
considered.  The source density is likely to be an underestimate since
parts of the maps have only been covered by a single chop at this time
and consequently have a higher noise level. However,
the uncorrected source
density is consistent with a surface density of point sources
more than sufficient for the science goals described in this paper.

\begin{figure*}
\centering { \includegraphics[width=8cm]{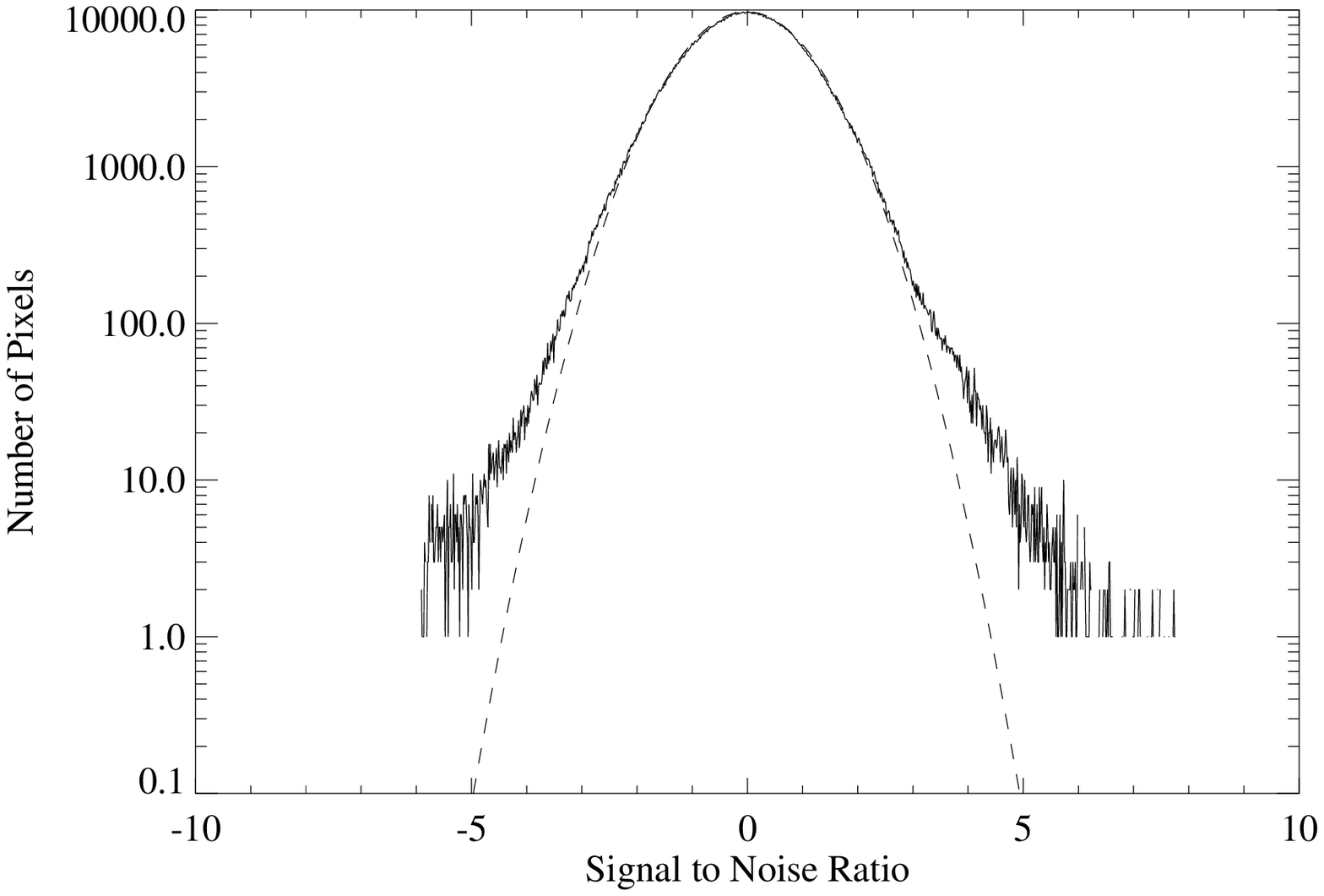}
\includegraphics[width=8cm]{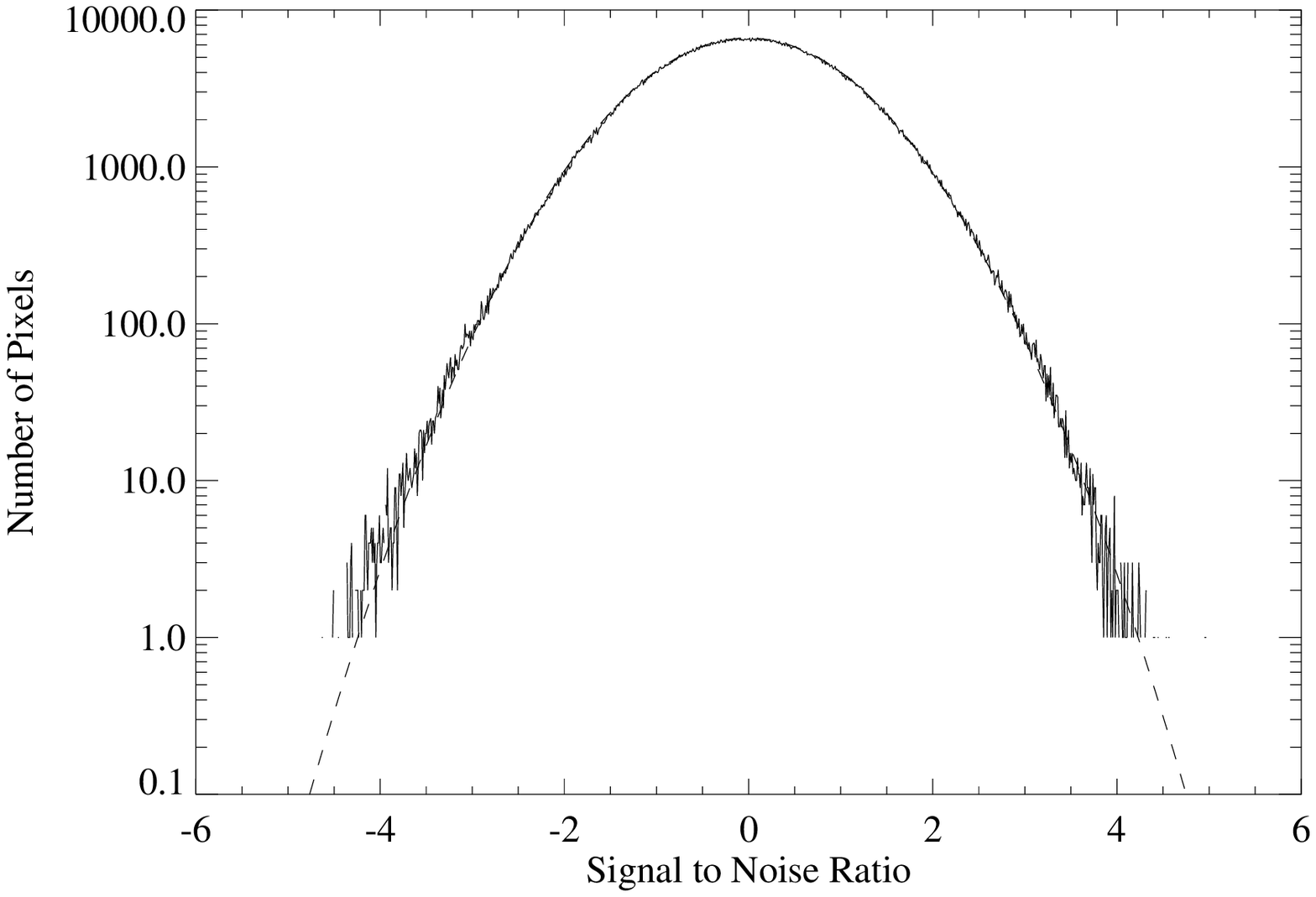}
\includegraphics[width=8cm]{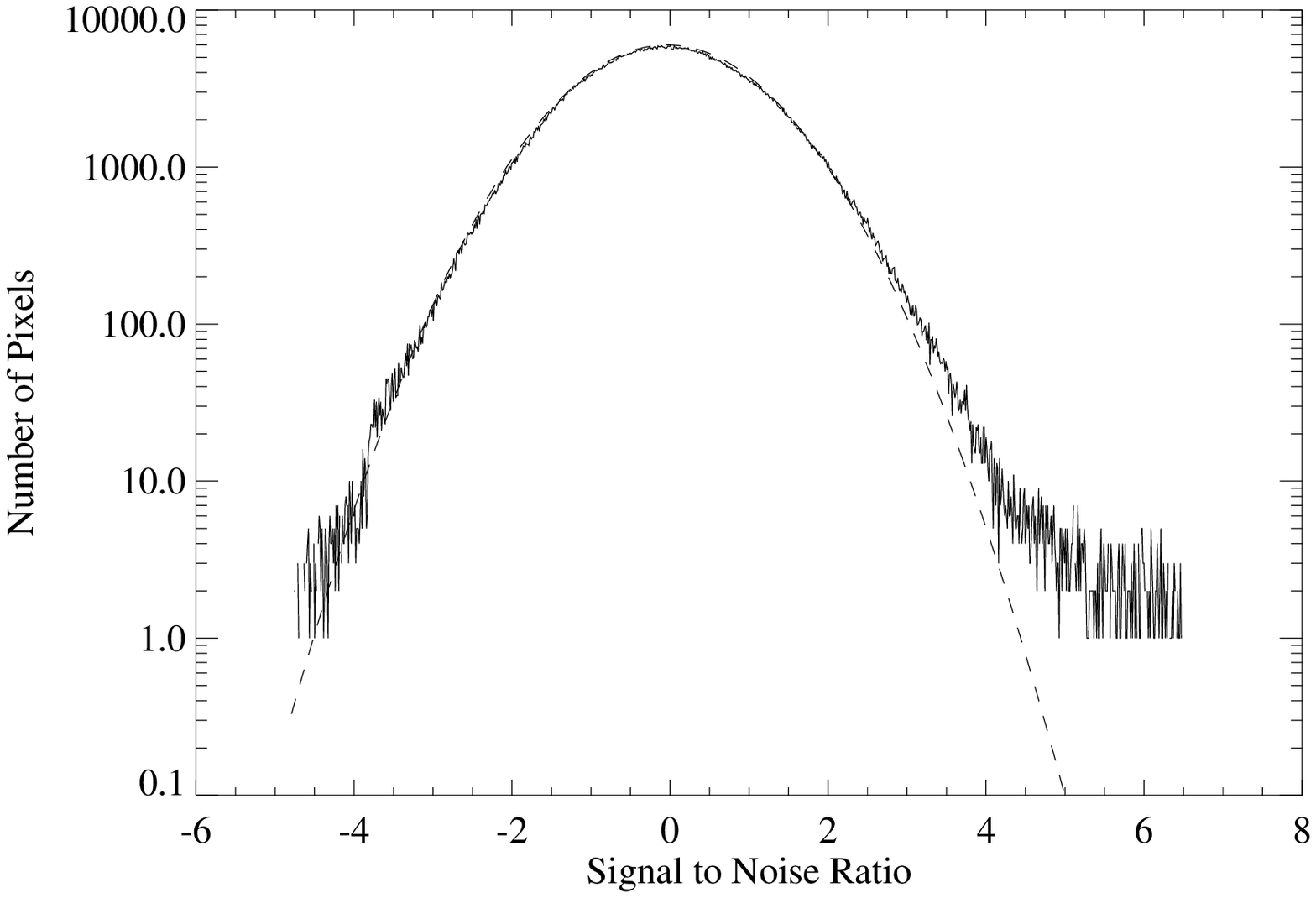}
\includegraphics[width=8cm]{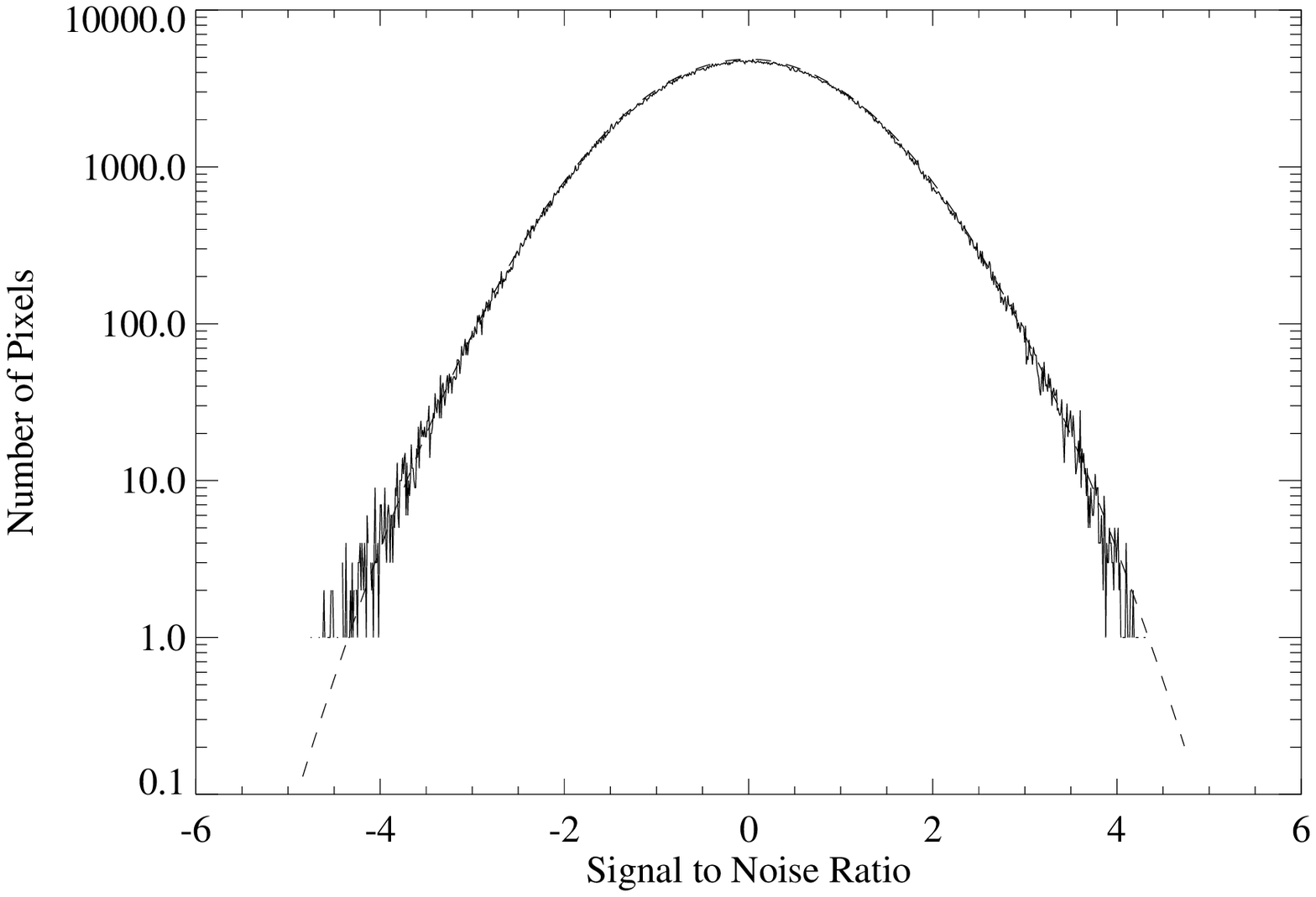}}
\caption {
\label{fig:snr_histogram}
S/N histograms of the point-source-filtered SHADES maps, with a
Gaussian fit to the data plotted as a dashed line.  The noise levels
are derived from the bolometer readouts using the
methodology described in Section \ref{sec:noise}, which are used in
the map making (Section \ref{sec:zerofootprint}) and source extraction
(Section \ref{sec:srcextr}) procedures.  
{\it Top:Left} Lockman Hole 850\,$\mu$m
map -- mean -0.0147, variance 1.067.  {\it Top:Right} Lockman
Hole 450\,$\mu$m map -- mean -0.00682, variance 1.020.  {\it
Bottom:Left} SXDF 850\,$\mu$m map -- mean -0.0374, variance
1.152.  {\it Bottom:Left} SXDF 450\,$\mu$m map -- mean
-0.002569, variance 1.111.  Note the very close fit to the data
at 450\,$\mu$m and the excess S/N at 850\,$\mu$m.}
\end{figure*}

Fig.\,\ref{fig:snr_histogram} shows the S/N histogram of the new maps.
A Gaussian can be fitted to the S/N data, though at high S/N levels an
excess above this fit shows there are real sources in the map.  This
is seen to be the case with the SXDF observations; in the Lockman
Hole, the effect of the combination of the deep strip (at one chop
throw only) and the noisier edges is evident from the higher excess of
pixels at higher S/N compared to the SXDF.  The Lockman Hole deep
strip has one chop throw only, so we also see an excess at large
negative S/N due to the side-lobes of the sources in this deep
strip. This negative excess is also present in the SXDF maps, but at a
much lower level because our chop strategy deliberately reduces the
chop holes (see Fig.\,\ref{fig:sims1}).  If there were no sources in
the map, the plot should follow the Gaussian shape. Instead, it is
possible to see the statistical detections at higher S/N as an excess
of pixels with that S/N, especially at 850\,$\mu$m.  Also, assuming
Gaussian random noise, we would expect to find
0.62\% of the survey beams to contain spurious detections at the
2.5\,$\sigma$
level.  This would mean that if there were no real sources in the map,
we would expect to see a total of about 60 spurious sources in the
Lockman Hole maps so far, and 44 spurious sources in 
the SXDF assuming a beam size of
14$\,^{\prime\prime}$ at 850\,$\mu$m.  Instead we detect approximately
270 source candidates at $\geq\,2.5\sigma$ in the Lockman Hole map, and
250 source candidates in the SXDF map. We have extended this analysis
to other $\sigma$ cuts in Table \ref{tab:numsnr}.

\begin{table}
\caption{\label{tab:numsnr} Numbers of sources at 850\,$\mu$m greater
than the S/N cut used in the source extraction procedure.  Note the
steep negative slope.  Numbers in brackets are the estimated
number of statistically 
spurious sources assuming Gaussian noise. } \centering
\begin{tabular}{cll}
\hline S/N & $N(>{\rm S/N})$ Lockman & $N(>{\rm S/N})$ SXDF\\ \hline
1.0&1773 (1530) & 1372 (1128) \\
2.0&648 (220) &524 (162) \\
3.0&98 (13) &106 (10) \\ 5&46 (2.25) &40 (1.66) \\
4.0&24 (0.3) &16 (0.2) \\
5.0&6 (0) &4 (0) \\
6.0&2 (0) &2 (0) \\ 5&1 (0) &1 (0) \\
\hline Median 1$\sigma$ noise& 2.28\,mJy & 2.14\,mJy \\ at 850\,$\mu$m\\
\hline
\end{tabular}
\end{table}

\subsection{Comparison with the 8-mJy survey}\label{sec:Comp}

A re-analysis of the 8-mJy Survey Lockman Hole East data was
carried out in order to test the new pipeline. Of the 21 published
sources at greater than 3.5\,$\sigma$ \citep{Scott02}, 4 
(LE850.9, 10, 15, 20) were rejected by \citet{Ivison02} because
they have $\sigma_{850}>3$\,mJy. This leaves 17. Of these, 12 were
found using the new analysis method.  
Of the 12 8-mJy survey sources with S/N greater than 3.5 that have
been reproduced, two lack a radio ID (LE850.4, LE850.11).  The 
possible implications of this were 
outlined in Section \ref{sec:crossid}.  For those sources
reproduced, Table \ref{tab:newsnr} shows the S/N of the source in the
new reduction. For those sources not reproduced, the S/N at the
position of the original 8-mJy Survey `detection' is shown. 

Of the sources previously detected between 3--3.5\,$\sigma$, only two
have been confirmed as peaks in the S/N distribution, and one 
source has positive flux at the position with S/N of greater than 3.
Extra chop data in the area have confirmed
one source at a significance level greater than 3.5\,$\sigma$ (LE850.29).

Only one of the nine missing sources originally found 
at a significance level greater than 3.5\,$\sigma$ have a
robust radio identification. This is source LE850.6,
which has been resolved into two sources, but neither new peak 
lies within 7$^{\prime\prime}$ of the original source position. 
Note that further analysis of the
\citet{Ivison02} radio identifications in the Lockman Hole was carried out by
\citet{Greve} and their results are used here.  Source LE850.13 has
been lost using this data-reduction method and does not return when
all six chop data are added to the region. 

One possibility is that the missing \citet{Scott02} sources were
spurious, perhaps due to weak bolometer noise spikes that escaped
clipping in the earlier analysis. However, this cannot be the whole
explanation since there is a weak positive signal
in the positions of the missing sources \citep[as quoted by][see Table
\ref{tab:newsnr}]{Scott02} in our refined re-analysis.

\begin{figure}
\centering { \includegraphics[width=8cm]{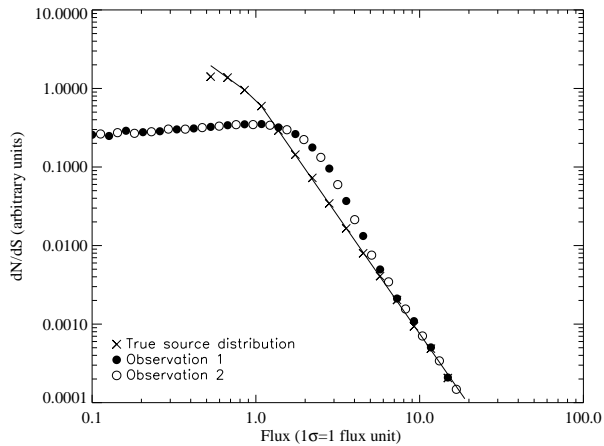}}
\caption {
\label{fig:eddingtonplot}
A simulation of Eddington bias for a sample of 10,000 sources with one
$\sigma$ equal to one flux unit.  Although flux boosting does occur
for a fraction of the simulated sources, the actual sources detected
between 3--3.5$\sigma$ varies between individual observations because
of the random effect of the noise.  Of the 10,000 sources simulated here,
$\sim$2600 were found with S/N between 3--
3.5$\sigma$ in the first and/or second observations,
but only 535 were found with S/N in the range 3--3.5$\sigma$ in the
first observation and $>3\sigma$ in the second.  Therefore, when
applying a significance threshold, the lower-significance source lists
derived from each observation are markedly different, despite the fact
that both observations yield clearly consistent Eddington-biased
source counts.}
\end{figure}
\renewcommand{\thefootnote}{\fnsymbol{footnote}}
\begin{table}
\caption{ 
\label{tab:newsnr} 
The S/N of the sources using the new data reduction software, applied
to the 8-mJy survey data only.  The S/N quoted is of the source that
corresponds to the old source (Less than 7$\,^{\prime\prime}$ from the
original 8-mJy survey source). Radio identifications
are from \citet{Greve} and
selected such that the probability that an apparent radio source
identification could be the result of chance is $p < 0.05$.  Of the
nine sources in the original ${\rm S/N}\geq 3.5$ list that have now
been lost, 4 have already been rejected by Ivison et al. (2002) as
likely reduction artifacts or seriously flux-boosted sources extracted
from the high-noise regions of the original 8-mJy maps (LE850.9, 10,
15 and 20, marked by an asterisk). Source LE850.6 has been
resolved into two blended sources, but does not strictly have an peak
within 7$\,^{\prime\prime}$ of the original source.  The new sources
found from our reduction are named with the prefix LE850.M, and do not
appear in the current consensus lists. Not listed here are the eight
sources found by our algorithm that {\it do} appear in the consensus
reduction.
}
\centering{
\begin{tabular}{llll}
\hline 8-mJy source & 8-mJy S/N & New S/N & Radio ID?\\
 \hline 
LE850.1 & 8.1 & 8.0 & Yes\\ 
LE850.2 & 5.2 & 5.1 & Yes\\ 
LE850.3 & 5.1 & 3.9 & Yes \\ 
LE850.4 & 5.0 & 5.1 & No\\ 
LE850.5 & 4.6 & 2.0 & No\\ 
LE850.6 & 4.5 & 2.8 & Yes\\ 
LE850.7 & 4.5 & 4.0 & Yes\\ 
LE850.8 & 4.4 & 3.8 & Yes\\ 
LE850.9\footnotemark[1] & 4.2 & 0.8 & No\\ 
LE850.10\footnotemark[1]  & 4.2 & 1.5 & No\\
LE850.11\footnotemark[2]  & 4.1 & 4.3 & No\\ 
LE850.12 & 4.0 & 4.3 & Yes\\ 
LE850.13 & 3.7 & 1.4 & No\\ 
LE850.14 & 3.6 & 5.2 & Yes\\ 
LE850.15\footnotemark[1]  & 3.6 & 1.8 & No\\ 
LE850.16 & 3.6 & 4.3 & Yes\\ 
LE850.17 & 3.5 & 1.3 & No\\ 
LE850.18 & 3.5 & 4.2 & Yes\\ 
LE850.19 & 3.5 & 1.7 & No\\ 
LE850.20\footnotemark[1]  & 3.5 & 0.2 & No\\ 
LE850.21 & 3.5 & 4.0 & Yes\\
\hline
LE850.27 & 3.4 & 3.3 & N/A\\
LE850.29 & 3.3 & 4.7 & N/A\\
LE850.32 & 3.2 & 3.2 & N/A\\
\hline 
LE850.M1 & -- & 3.88 & Yes\\ 
LE850.M2 & -- & 3.77 & ?\\
LE850.M3 & -- & 3.73 & N/A\\ 
LE850.M4 & -- & 3.51 & N/A\\
\hline
\end{tabular}
\begin{tabular}{l}
\footnotemark[2] Source rejected in \citet{Ivison02} for \\
its high noise value but considered robust here\\
\hline
\end{tabular}}
\end{table}

One other possibility that could lead to {\it apparently\/} spurious
sources, in the sense that the sources seem to be non-repeatable, is
Eddington bias \citep{Eddington}. This bias, sometimes confused with
Malmquist bias \citep{Teerikorpi}, describes a feature of observing a
population of objects above a given flux limit with a negative-sloping
source count. Random errors in the flux measurements of the objects
can systematically alter the source counts measured above the flux
limit, such that more sources have their flux densities boosted above
the flux limit, than those that fall below the limit. This makes there
appear to be more sources close to the flux limit than in the true
population, the effect being more pronounced for lower S/N thresholds.
A simulation of this effect on 10,000 sources can be seen in Fig. 8.
When observations are repeated, different sources are boosted above
the flux limit while others again drop below the flux limit.  The
number of sources observed in each of the first and second
observations with a S/N in the range 3--3.5$\sigma$ is $\sim$2600 but
the number of sources with S/N in the range 3--3.5$\sigma$ in the
first observation and $>3\sigma$ in second is only 535.  In this way
it is possible to have up to 80\% of the 3--3.5$\sigma$ sources not
appearing in both samples.  This 80\% is made up of those sources that
are flux boosted in the first observation and not in the second, or
vice versa, and it will happen even if all the 3--3.5$\sigma$ lists
are 100\% reliable.  By re-analysing the noise in the 8-mJy survey
maps, we have re-weighted the observations, and effectively resampled
the noise, although the effect will probably be smaller than that
shown in Fig.  8 because the raw data are the same and it is the
dominant sky noise that has been re-measured.  This might explain why
the lower significance 3--3.5$\sigma$ source lists from the 8-mJy
survey data reduction and the SHADES data reduction contain only 3
sources in common.

Fifteen new $\geq 3.5\sigma$ sources have been identified using the
new source-extraction method in the 8-mJy survey area, of which only
seven are not also in the consensus lists. Three of these fifteen have
noise greater than 3\,mJy.  One of the new sources in the consensus
list was first reported in full detail in \citet{Serjeant04} with two
possible {\it Spitzer\/} detections which agree strongly with the
radio identifications presented there.  Although not detected at 450 and
1200\,$\mu$m, weak positive signals of 1.6$\sigma$ and 1.8$\sigma$ have
been found at this source position in both maps. The four sources {\it
not\/} in the consensus list and with noise less than 3\,mJy are listed
in Table \ref{tab:newsnr}.

A future paper 
will perform a detailed comparison of the pipelines over all the SHADES
area, i.e. not just restricted to the 8-mJy Lockman Hole field.

\section{Conclusions and survey progress}\label{sec:discussion}
To February 2004 the SCUBA Half Degree Extragalactic Survey areas covered a
region of 720\, arcmin$^{2}$ (approximately 40\% of the total expected
area) using 1843 individual jiggle-maps that have been observed over
139 nights during the period March 1998 -- February 2004.

Source extraction from these maps indicate at least $653\pm57$
sources deg$^{-2}$ having S/N $>3$ at 850\,$\mu$m in these survey fields
(uncorrected for Eddington bias), consistent with a surface density of
point sources more than sufficient for the science goals described in
this paper.

This paper has outlined the SHADES survey goals, data-taking and data
reduction strategies, as part of which a new SCUBA reduction pipeline
has been developed using IDL, based upon the reduction pipeline used
for the 8-mJy survey.  A test of this data-reduction and 
source-extraction method was made by comparing the new maps with the sources
extracted from the original 8-mJy survey Lockman Hole East maps that
form a sub-set of the SHADES maps. Of the 17 more secure ($\geq
3.5\sigma$) sources in the 8-mJy survey, 13 have been reconfirmed of
which 11 have radio identifications. 13 new candidate sources
with ${\rm S/N}\geq 3.5$ and ($\sigma_{850} <3.5$\,mJy) have been
identified using the new data-reduction method, nine of which appear
in the consensus between other reductions which will be discussed in
future papers.

A full presentation and analysis of the sub-mm source counts derived from
the interim SHADES maps discussed here will be presented in a separate
paper. This paper will also compare the results of the data reduction and
source extraction described here, with the outcome of three further
independent reductions/extractions carried out at IfA Edinburgh, INAOE
Mexico, and UBC, Vancouver.

The SHADES consortium maintains a public web page at
\url{http://www.roe.ac.uk/ifa/shades/}.

\section*{Acknowledgements}
Major parts of this work were supported by PPARC and by the Canadian
NSERC. This work is also supported by Nuffield Foundation Grant number
NAL/00529/G.  AJB, ACE, OA, PNB and IRS acknowledge the Royal Society
for generous funding.  DHH, IA and JW are supported by CONACYT grants
39953-F and 39548-F.  CPP acknowledges a European Union fellowship to
Japan.  EvK acknowledges funding from by the Austrian Science
Foundation FWF under grant P15868.  The JCMT is operated by the Joint
Astronomy Centre on behalf of the UK Particle Physics and Astronomy
Research Council, the Canadian National Research Council and the
Netherland Organization for Scientific Research.

\setlength{\bibhang}{2.0em}

\end{document}